\documentclass[aps,prd,preprint,superscriptaddress,showpacs,preprintnumbers,amsmath,amssymb]{revtex4-1}

\usepackage{graphicx}
\usepackage{epstopdf}
\usepackage{dcolumn}
\usepackage{float}
\usepackage[normalem]{ulem}
\usepackage{color}

\newcommand{\be}{\begin{equation}}
\newcommand{\ee}{\end{equation}}
\newcommand{\bea}{\begin{eqnarray}}
\newcommand{\eea}{\end{eqnarray}}

\newcommand{\bfk}{\mbox{\boldmath $k$}}
\newcommand{\bfq}{\mbox{\boldmath $q$}}

\def\lsim{\mathrel{\rlap{\lower4pt\hbox{\hskip1pt$\sim$}}\raise1pt\hbox{$<$}}}
\def\gsim{\mathrel{\rlap{\lower4pt\hbox{\hskip1pt$\sim$}}\raise1pt\hbox{$>$}}}

\begin{document}

\title{Transverse Single Spin Asymmetry in 
{\boldmath$e+p^\uparrow \to  e+J/\psi +X $ and $Q^2$ -evolution of Sivers Function-II\\}}

\author{Rohini M. Godbole}
\email{rohini@cts.iisc.ernet.in}
\author{Abhiram Kaushik}
\email{abhiramb@cts.iisc.ernet.in}
\affiliation{Centre for High Energy Physics, Indian Institute of Science, Bangalore, India.}

\author{Anuradha Misra}
\email{misra@physics.mu.ac.in}
\author{Vaibhav S. Rawoot}
\affiliation{Department of Physics, University of Mumbai, Mumbai, India.}
\email{vaibhavrawoot@gmail.com}

\date{\today}
\begin{abstract}
We present estimates of Single Spin Asymmetry (SSA) in the electroproduction of $J/\psi$ taking into account the  transverse momentum dependent (TMD) evolution 
of the gluon Sivers function.  We estimate SSA for JLab, HERMES, COMPASS and eRHIC energies using color evaporation model of 
$J/\psi$. We have calculated the asymmetry using recent parameters extracted by Echevarria \emph{et al.}
using the CSS approach to TMD evolution. These recent TMD evolution fits are based on the 
evolution kernel in which the perturbative part is  resummed up to  next-to-leading logarithms (NLL) accuracy. We have also estimated the asymmetry by using parameters which had been obtained by a fit by Anselmino \emph{et al.}, using both an exact numerical and an approximate analytical solution of the TMD evolution equations. We find that the variation among the different estimates obtained using TMD evolution is much smaller than between these on one hand and the estimates obtained using DGLAP evolution on the other. Even though the use of TMD evolution causes an overall reduction in asymmetries compared to the ones obtained without it, they remain sizable. Overall, upon use of TMD evolution, predictions for asymmetries stabilize.
\end{abstract}
 
\maketitle

\section{\label{intro}Introduction}
There has been a lot of work done lately on the phenomenology of TMDs (Transverse Momentum Dependent PDFs, Fragmentation Functions etc), their measurement and $Q^2$ evolution~\cite{Aybat:2011zv,Aybat:2011ge}. TMDs give details of the intrinsic transverse momenta of partons, providing an understanding of the 3-dimensional structure of nucleons. Their $Q^2$ evolution has non-perturbative contributions as opposed to the case of collinear distributions, whose evolution is completely perturbative.  A knowledge of the TMDs can be obtained by using the Single Spin Asymmetries (SSAs) observed in scattering experiments involving a single transversely polarized hadron~\cite{tmd-fact1, Sivers1990, tmd-fact2}. These include Drell-Yan (DY) scattering and semi-inclusive Deep Inelastic Scattering (SIDIS). One of the ways to analyse these SSAs is based on  a Transverse Momentum Dependent factorization scheme. Such a scheme was first provided by Collins and Soper~\cite{fact,Collins:2011book} which  has, since then, been used to study the above processes. In TMD factorization, the transverse momenta of the partons are not integrated over, as they are in the standard colinear factorization schemes of QCD.

One of the most important SSAs is the Sivers Asymmetry. It is due to a TMD called the Sivers function, which gives the probability of finding an unpolarized parton inside a transversely polarized nucleon. Many fits of the Sivers function are available which are extracted from experimental data of SIDIS from the HERMES, COMPASS and JLAB experiments~\cite{Anselmino-PRD72,kp09}. The most recent fit, which we use in this paper, is by Echevarria \emph{et al.}~\cite{Echevarria:2014xaa} who have fitted the Sivers function in the Torino parametrization, one of the two commonly used parametrizations of the Siver function, the other being the Bochum parametrisation \cite{Aybat:2011ge}.We have exclusively used the Torino parametrisation in this as well as our previous works on the subject.  Echevarria \emph{et al.} have performed a global fit to all the experimental data on SIDIS from HERMES \cite{hermes09}, COMPASS~\cite{Alekseev:2008aa,Adolph:2012sp} and JLAB~\cite{Qian:2011py} experiments. They have used the Sivers function fits so obtained to make predictions for the Sivers asymmetry in Drell Yan and $W^+$ and $W^-$ boson production. These predictions agree well with data, giving a $\chi^2/d.o.f$ $\approx$ $1.3$. Previous fits given by Anselmino \emph{et al.}~\cite{Anselmino:2012aa} had extracted the Sivers function in SIDIS data from the HERMES~\cite{hermes09} and COMPASS~\cite{Bradamante:2011xu} experiments alone. The earliest fits of the Sivers function had assumed that the transverse momentum behaviour factorized from the collinear distributions and did not evolve. This may be a reasonable approximation to make for low $Q^2$ processes but is not valid for high $Q^2$ processes as was the case for COMPASS data. 

Both fits by Echevarria \emph{et al.} and Anselmino \emph{et al.}, used here, incorporate the evolution of the transverse momentum dependence. The Echavarria \emph{et al.} fits differ from the Anselmino \emph{et al.} fits in two aspects: Firstly, they use a certain prescription for the initial scale of the evolution kernel explained in Ref. ~\cite{Echevarria:2014xaa} in order to simplify the evolution equations and secondly, their kernel is consistently resummed to next-to-leading-logarithmic (NLL) accuracy. 

  In the present work, we make predictions for Sivers asymmetry in low virtuality electroproduction of $J/\psi$, using the latest fits by Echevarria \emph{et al.}~\cite{Echevarria:2014xaa}. The present work deals with the gluon Sivers asymmetry in the leptoproduction of charmonium. There are not enough data to fit the gluon Sivers function directly. Therefore, we parametrize it in terms of the quark Sivers function~\cite{Boer-PRD69(2004)094025}, to give estimates of asymmetry in leptoproduction of $J/\psi$. 

The leptoproduction of heavy flavours (open and closed) in general, and of charmonium, in particular, is a direct probe of the gluon content of the proton. At leading order, this involves a gluon and a photon fusing to form a $c\bar c$ pair. It was first studied in this context in Ref.~\cite{gr78}. In our earlier work we had extended the idea to the case of the gluon Sivers function~\cite{Godbole:2012bx, Godbole:2013bca}.  We had made predictions for Sivers asymmetry in the process $e+p^\uparrow \rightarrow e +J/\psi + X$, first using DGLAP evolved TMDs~\cite{Godbole:2012bx} and later, PDFs and Sivers function taking evolution of the transverse momentum distribution into account.~\cite{Godbole:2013bca}. In this paper, we would like to assess the dependence of asymmetries on the different aspects of the implementation of the QCD evolution of TMDs. The estimates in Ref. \cite{Godbole:2013bca} were based on a formalism given by Anselmino \emph{et al.} where they had used an  analytical solution of an approximate form of the TMD evolution equation~\cite{Anselmino:2012aa}. In the current work, we use the exact treatment of TMD evolution. Secondly, we also use a new parametrization of the Sivers function obtained from fits performed with a NLL resummed evolution kernel \cite{Echevarria:2014xaa}. We give revised estimates of the Sivers Asymmetry in charmonium leptoproduction.

In this work, the Colour Evaporation Model(CEM) is used to get the cross-section for the production of $J/\psi$. The details of the Color Evaporation model can be found in Ref.~\cite{hal,fri}. In section II, a brief summary of the  construction of the Sivers asymmetry observables is given. Section III deals with the evolution of the TMDs. In section IV, we give the form of the TMDs with the evolution kernel resummed at NLL. Section V gives the details of the fits that we have used. This is followed by a summary and analysis of the results in Section VI.

\section{SINGLE SPIN ASYMMETRY IN $J/\psi$ PRODUCTION : FORMALISM}
In the Colour Evaporation Model(CEM), the leading order (LO) cross-section for production of $J/\psi$ is proportional to the rate of $c\bar{c}$ production integrated over the invariant mass squared of the $c\bar{c}$ pair ranging from $4m_c^2$ to $4m_D^2$ where $m_D$ is the open charm production threshold~\cite{cem0}
\be
\sigma^{ep\rightarrow e+J/\psi+X}=
\int_{4m_c^2}^{4m_D^2} dM^2_{c\bar{c}} \int dx_\gamma \> dx_g\> f_{\gamma/e} (x_\gamma )\> f_{g/p}(x_g) 
\>\frac{d\hat{\sigma}^{\gamma g\rightarrow c\bar{c}}}{dM_{c\bar{c}}^2}.
\label{xsec-gammap}
\ee
Here, $f_{g/p}(x)$ is the gluon PDF and $f_{\gamma/e}(\xi)$ is the well-known Weiszacker-Williams function~\cite{wwf1,wwf2}.

The SSA in the scattering of electrons off a transversely polarized proton target arises due to transverse momenta of the partons, so we use a generalized expression that takes into account the transverse momentum behaviour of the Weizsacker-Williams function and the gluon PDF~\cite{Godbole:2012bx}
\be
\frac{d\sigma^{e+p^\uparrow\rightarrow e+J/\psi + X}}{d M^2}=
\int  dx_\gamma\> dx_g\> [d^2\bfk_{\perp\gamma}d^2\bfk_{\perp g}]\>
f_{g/p^{\uparrow}}(x_{g},\bfk_{\perp g})
f_{\gamma/e}(x_{\gamma},\bfk_{\perp\gamma})
\frac{d\hat{\sigma}^{\gamma g\rightarrow c\bar{c}}}{dM^2}
\label{dxec-ep}
\ee
where $M^2\equiv M_{c\bar{c}}^2$. The gluon Sivers function gives the difference between $d\sigma^\uparrow$ and $d\sigma^\downarrow$
\be
d\sigma^\uparrow-d\sigma^\downarrow=
\int dx_\gamma\> dx_g\> d^2\bfk_{\perp\gamma}\>d^2\bfk_{\perp g}\>
\Delta^{N}f_{g/p^{\uparrow}}(x_{g},\bfk_{\perp g})\>
f_{\gamma/e}(x_{\gamma},\bfk_{\perp\gamma})\>
{d\hat{\sigma}}^{\gamma g\rightarrow c\bar{c}}
\label{dxsec-nssa}
\ee
The numerator and denominator of the asymmetry are given by ~\cite{Godbole:2012bx} 
\bea
\frac{d^{4}\sigma^\uparrow}{dydM^2d^2\bfq_T}-\frac{d^4\sigma^\downarrow}{dydM^2d^2\bfq_T}=
\frac{1}{s}\int [d^2\bfk_{\perp\gamma}d^2\bfk_{\perp g}]
\Delta^{N}f_{g/p^{\uparrow}}(x_{g},\bfk_{\perp g})
f_{\gamma/e}(x_{\gamma},\bfk_{\perp\gamma}) \nonumber\\
\times\>\delta^2(\bfk_{\perp\gamma}+\bfk_{\perp g}-\bfq_T)
\hat\sigma_{0}^{\gamma g\rightarrow c\bar{c}}(M^2)
\label{num-ssa}
\eea

and 

\bea
\frac{d^{4}\sigma^\uparrow}{dydM^2d^2\bfq_T}+\frac{d^4\sigma^\downarrow}{dydM^2d^2\bfq_T}=
\frac{2}{s}\int [d^2\bfk_{\perp\gamma}d^{2}\bfk_{\perp g}]
f_{g/p}(x_g,\bfk_{\perp g})
f_{\gamma/e}(x_{\gamma},\bfk_{\perp\gamma}) \nonumber\\
\times\>\delta^2(\bfk_{\perp\gamma}+\bfk_{\perp g}-\bfq_T)
\hat\sigma_{0}^{\gamma g\rightarrow c\bar{c}}(M^2)
\label{den-ssa}
\eea

where 
\be
x_{g,\gamma} = \frac{M}{\sqrt s} \, e^{\pm y}
\label{x-gammag}
\ee
with the partonic cross section given by~\cite{gr78}
\be
\hat{\sigma_0}^{\gamma g\rightarrow c\bar{c}}(M^2)=
\frac{1}{2}e_{c}^2\frac{4\pi\alpha\alpha_s}{M^2}
[(1+ v -\frac{1}{2} v^2)\ln{\frac{1+\sqrt{1- v}}{1-\sqrt{1-v}}}
-(1+v)\sqrt{1-v}].
\label{jpsics}
\ee
Here, $\displaystyle v=\frac{4 m_c^2}{M^2}$ and $M^2\equiv\hat{s}$.
We integrate Eqs. (\ref{num-ssa})  and (\ref{den-ssa}) over $M^2$, 
to obtain  the difference and sum of $\displaystyle\frac{d^{3}\sigma^\uparrow}{dyd^2\bfq_T}$ 
and $\displaystyle\frac{d^3\sigma^\downarrow}{dyd^2\bfq_T} $for $J/\psi$ production. \\
[3ex]The sum and difference of the differential cross-sections with respect to $y$ is then

\bea
\frac{d\sigma ^\uparrow}{dy} \, - \, \frac{d\sigma ^\downarrow}{dy}=\int d\phi_{q_T}\int q_T\>dq_T
\int_{4m^2_c}^{4m^2_D}[dM^{2}]\int[d^2\bfk_{\perp g}]
\Delta^{N}f_{g/p^{\uparrow}}(x_{g},\bfk_{\perp g}) \nonumber \\
\times\>f_{\gamma/e}(x_{\gamma},\bfq_T-\bfk_{\perp g}) \>
\hat\sigma_{0}(M^2)\>\sin({\phi}_{q_T}-\phi_S) \label{dxsec-y}
\eea
and
\bea
\frac{d\sigma ^\uparrow}{dy} \, + \, \frac{d\sigma ^\downarrow}{dy}= 
2\int d\phi_{q_T}\int q_T\>dq_T\int_{4m^2_c}^{4m^2_D}[dM^{2}]\int[d^{2}\bfk_{\perp g}]
f_{g/p}(x_g,\bfk_{\perp g}) \nonumber \\
\times\>f_{\gamma/e}(x_{\gamma},\bfq_T-\bfk_{\perp g})\,
\hat{\sigma}_0(M^2). 
\label{txsec-y}
\eea
and the sum and difference of the differential cross-sections with respect to $q_T$ is 
\bea
\frac{d\sigma ^\uparrow}{dq_T} \, - \, \frac{d\sigma ^\downarrow}{dq_T}=\int d\phi_{q_T}\int q_T\>dy
\int_{4m^2_c}^{4m^2_D}[dM^{2}]\int[d^2\bfk_{\perp g}]
\Delta^{N}f_{g/p^{\uparrow}}(x_{g},\bfk_{\perp g}) \nonumber \\
\times\>f_{\gamma/e}(x_{\gamma},\bfq_T-\bfk_{\perp g}) \>
\hat\sigma_{0}(M^2)\>\sin({\phi}_{q_T}-\phi_S) \label{dxsec-qT}
\eea
and
\bea
\frac{d\sigma ^\uparrow}{dq_T} \, + \, \frac{d\sigma ^\downarrow}{dq_T}= 
2\int d\phi_{q_T}\int q_T\>dy\int_{4m^2_c}^{4m^2_D}[dM^{2}]\int[d^{2}\bfk_{\perp g}]
f_{g/p}(x_g,\bfk_{\perp g}) \nonumber \\
\times\>f_{\gamma/e}(x_{\gamma},\bfq_T-\bfk_{\perp g})\,
\hat{\sigma}_0(M^2). 
\label{txsec-qT}
\eea

The weighted Sivers asymmetry is defined as~\cite{vogelsang-weight}
\be
A_N^{\sin({\phi}_{q_T}-\phi_S)} =\frac{\int d\phi_{q_T}
[d\sigma ^\uparrow \, - \, d\sigma ^\downarrow]\sin({\phi}_{q_T}-\phi_S)}
{\int d{\phi}_{q_T}[d{\sigma}^{\uparrow} \, + \, d{\sigma}^{\downarrow}]}
\label{weight-ssa}
 \ee
where $d\sigma^\uparrow$ is differential cross section in $q_T$ or y variable, and 
${\phi}_{q_T} $ and $\phi_S$ are the azimuthal angles of the $J/\psi$ and proton spin respectively.  For the asymmetry with respect to $y$, we use Eqs. (\ref{dxsec-y}) and (\ref{txsec-y}) in Eqn. (\ref{weight-ssa}). For calculating the asymmetry with respect to $q_T$ we use (\ref{dxsec-qT}) and (\ref{txsec-qT}).

The transverse momentum dependence of the Weiszacker-Williams function is taken to be gaussian:
\be
f_{\gamma/e}(x_\gamma,k_{\bot \gamma})=f_{\gamma/e}(x_\gamma)\frac{1}{\pi\langle k^{2}_{\bot \gamma}\rangle} 
e^{-k^{2}_{\bot \gamma}/\langle{k^{2}_{\bot \gamma}\rangle}}.
\label{gauss-g}
\ee 

The Sivers function and the transverse momentum dependent form of the PDF are given in Section IV.

\section{$Q^2$-Evolution of TMD}

In this section, we present a brief outline of the energy evolution of transvese momentum dependent functions as given in Ref.~\cite{Echevarria:2014xaa}.
A general transverse momentum dependent distribution (TMD) $F(x,k_\perp;Q)$ can be expressed in a 2-dimensional coordinate space (called $b$-space) by a fourier transform as
\be
F(x,b;Q)=\int d^2k_\perp e^{-ik_\perp.b}F(x,k_\perp;Q)
\ee
We will work with the $b$-space TMDs as the energy evolution is more naturally described in b-space. It is given by 
\be
F(x,b,Q_f)=F(x,b,Q_i)R_{pert}(Q_f,Q_i,b_*)R_{NP}(Q_f,Q_i,b)
\label{evolution}
\ee
where $R_{pert}$ is the perturbative part of the evolution kernel, $R_{NP}$ is the nonperturbative part of kernel and $b_*=b/\sqrt{1+(b/b_\text{max})^2}$.

The perturbative part is given by
\be
R(Q_f,Q_i,b)=\exp\left\{-\int_{Q_i}^{Q_f}\frac{d\mu}{\mu}\left(A\ln\frac{Q_f^2}{\mu^2}+B\right)\right\}\left(\frac{Q_f^2}{Q_i^2}\right)^{-D(b;Q_i)}
\ee
where A=$\Gamma_\text{cusp}$ and B=$\gamma^V$, with $\frac{dD}{d\ln\mu}=\Gamma_\text{cusp}$. The anomalous dimensions $\Gamma_{\text{cusp}}$ and $\gamma^V$ are known upto three loop level~\cite{Idilbi:2006dg}. 

The nonperturbative exponential part, called the Sudakov factor, is fixed by fits to data. It contains a $Q$-dependent factor universal to all TMDs and another factor which gives the gaussian width in $b$-space of the particular TMD.
\be
R_{NP}=\exp\left\{-b^2\left(g_1^\text{TMD}+\frac{g_2}{2}\ln\frac{Q_f}{Q_i}\right)\right\}
\ee

The $b_*=b/\sqrt{1+(b/b_\text{max})^2}$ prescription used in Eq.(\ref{evolution}) is used to stitch together the perturbative part(which is valid at small $b$) and nonperturbative part(which is valid at large $b$). When $b\ll b_\text{max}$, $b_*\approx b$, whereas at higher values of $b$, $b_*\approx b_\text{max}$. As shown in Ref.~\cite{Echevarria:2014xaa}, for consistency upto NLL, expanding the TMD $F(x,b;Q)$ at the initial scale in terms of its corresponding collinear function and keeping only the LO term, which is just the collinear PDF, we finally get
\begin{align}
f_{q/H}(x,b;Q_f)=&f_{q/H}(x,Q_i)\exp\left\{-\int_{Q_i}^{Q_f}\frac{d\mu}{\mu}\left(A\ln\frac{Q^2}{\mu^2}+B\right)\right\}\left(\frac{Q_f^2}{Q_i^2}\right)^{-D(b^*;Q_i)} \nonumber \\ \times&\exp \left\{-b^2\left(g_1^\text{pdf}+\frac{g_2}{2}\ln\frac{Q_f}{Q_i}\right)\right\}
\end{align}

The CSS evolution of the Sivers function is discussed in Ref.~\cite{Aybat:2011ge}. It has been shown that the derivative of the Sivers function in $b$-space,
\be
f'^{\perp}_{1T}(x,b;\mu)\equiv\frac{\partial f^{\perp}_{1T}(x,b;\mu)}{\partial b}
\ee
satisfies the same evolution equation as the unpolarized TMDPDF. Therefore, the evolution equation is given by
\begin{align}
f'^{\perp g}_{1T}(x,b;Q_f)=&f'^{\perp g}_{1T}(x,b;Q_i)\exp\left\{-\int_{Q_i}^{Q_f}\frac{d\mu}{\mu}\left(A\ln\frac{Q^2}{\mu^2}+B\right)\right\}\left(\frac{Q_f^2}{Q_i^2}\right)^{-D(b^*;Q_i)} \nonumber\\ 
\times&\exp \left\{-b^2\left(g_1^\text{sivers}+\frac{g_2}{2}\ln\frac{Q_f}{Q_i}\right)\right\}
\end{align}
\\
Now the azimuth dependent part of the Sivers function (in $b$-space) is~\cite{Aybat:2011ge}
\be
f^{\perp q(\alpha)}_{1T}(x,b)=\frac{1}{M_p}\int d^2k_T e^{-i\bf k_T.\bf b}k^\alpha_Tf^{\perp q}_{1T}(x,k_T^2)
\ee
Expanding this in $b$, we get
\begin{align}
f^{\perp q(\alpha)}_{1T}(x,b)=&\frac{1}{M_p}\int d^2k_T\left[1-ik^\beta_Tb^\beta+...\right]k^\alpha_Tf^{\perp q}_{1T}\nonumber (x,k_T^2)\\=&-\frac{ib^\alpha}{2M_p}\int d^2k_T |k_T|^2f^{\perp q}_{1T}(x,k_T^2)+...\nonumber
\\=&\frac{ib^\alpha}{2}T_{q,F}(x,x)+...
\end{align}
Here, $T_{q,F}$ is the twist-3 Qiu-Sterman quark-gluon correlation function. It is the first $k_T$-moment of the quark Sivers function~\cite{Boer:2003cm,Kang:2011hk}. This equation was obtained in Ref.~\cite{Kang:2011mr}.

Now using the relation between the azimuthal part of the Sivers function and the derivative of the Sivers function, we have~\cite{Aybat:2011ge}
\begin{align}
f'^{\perp}_{1T}(x,b)=&-i\frac{M_pb}{b^\alpha}f^{\perp q(\alpha)}_{1T}(x,b)\nonumber \\
\simeq& \frac{M_pb}{2}T_{q,F}(x,x)
\end{align}

Therefore for the Sivers function, we finally get~\cite{Echevarria:2014xaa},
\begin{align}
f'^{\perp g}_{1T}(x,b;Q_f)=&\frac{M_p b}{2}T_{g,F}(x,x,Q_i)\exp\left\{-\int_{Q_i}^{Q_f}\frac{d\mu}{\mu}\left(A\ln\frac{Q^2}{\mu^2}+B\right)\right\}\left(\frac{Q_f^2}{Q_i^2}\right)^{-D(b^*;Q_i)} \nonumber\\ 
\times&\exp \left\{-b^2\left(g_1^\text{sivers}+\frac{g_2}{2}\ln\frac{Q_f}{Q_i}\right)\right\}
\end{align}

\section{CSS Evolution at NLL}
$A$ and $B$ in Eqs. (18) and (24) are $\Gamma_{\text{cusp}}$ and $\gamma^V$ respectively and can be expanded perturbatively. As mentioned earlier, the expansion coefficients are known upto 3-loop level. The $D$ term can also be expanded perturbatively as $D=\sum^\infty_{n=1}D^{(n)}(\alpha_s/n)^n$.
The expansion coefficients with the appropriate gluon anomalous dimensions at
NLL are~\cite{Echevarria:2014xaa}
\begin{align}
A^{(1)}&=C_A
\\
A^{(2)}&=\frac{1}{2}C_F\left(C_A\left(\frac{67}{18}-\frac{\pi^2}{6}\right)-\frac{5}{9}C_AN_f\right)
\\
B^{(1)}&=-\frac{1}{2}\left(\frac{11}{3}C_A-\frac{2}{3}N_f\right)\\
D^{(1)}&=\frac{C_A}{2}\ln\frac{Q_i^2b^{*2}}{c^2}
\end{align}

Choosing the initial scale $Q_i=c/b^*$, the $D$ term vanishes at NLL.
The expressions for the TMDs therefore become
\begin{align}
f_{g/p}(x,b;Q)=&f_{g/p}(x,c/b^*)\exp\left\{-\int_{c/b^*}^{Q}\frac{d\mu}{\mu}\left(A\ln\frac{Q^2}{\mu^2}+B\right)\right\} \nonumber \\
&\times \exp \left\{-b^2\left(g_1^\text{pdf}+\frac{g_2}{2}\ln\frac{Q b^*}{c}\right)\right\}
\end{align}

\begin{align}
f'^{\perp g}_{1T}(x,b;Q)=&\frac{M_p b}{2}T_{g,F}(x,x,c/b^*)\exp\left\{-\int_{c/b^*}^{Q}\frac{d\mu}{\mu}\left(A\ln\frac{Q^2}{\mu^2}+B\right)\right\} \nonumber\\ 
\times&\exp \left\{-b^2\left(g_1^\text{sivers}+\frac{g_2}{2}\ln\frac{Qb^*}{c}\right)\right\}
\label{SiversExpression}
\end{align} 

This is related to the Sivers function given in Eqn. (\ref{dxsec-nssa}) by
\be
\Delta^{N}f_{g/p^{\uparrow}}(x_{g},\bfk_{\perp g},Q)=-2\frac{k_{\perp g}}{M_p}f^{\perp g}_{1T}(x_g,k_{\perp g};Q)\cos\phi_{k_\perp}
\ee
where $f^{\perp g}_{1T}(x_g,k_{\perp g};Q)$ and similarly $f(x,k_{\perp g},Q)$ can be obtained from $f'^{\perp g}_{1T}(x_g,b;Q)$ and $f(x,b;Q)$ by doing a fourier transform as shown in Ref.~\cite{Anselmino:2012aa}
\be
f_{g/p}(x_g,k_\perp;Q)=\frac{1}{2\pi}\int^\infty_0db\text{ }bJ_0(k_\perp b)f_{g/p}(x_g,b;Q)
\ee
\be
f^{\perp g}_{1T}(x_g,k_\perp;Q)=-\frac{1}{2\pi k_\perp}\int^\infty_0db\text{ }bJ_1(k_\perp b)f'^{\perp g}_{1T}(x_g,b;Q)
\ee

\section{Numerical estimates}
We adopt the formalism of Ref.~\cite{Echevarria:2014xaa} and assume that the Qiu-Sterman function is proportional to the unpolarized collinear PDFs~\cite{Kouvaris:2006zy, Echevarria:2014xaa}.
\begin{align}
T_{q,F}(x,x,Q)=\mathcal{N}_q(x)f_{q/P}(x,Q)
\end{align}
where for $\mathcal{N}_q(x)$, as in our previous work, we use the Torino parametrization.
\be 
{\mathcal N}_f(x) = N_f x^{a_f} (1-x)^{b_f} \frac{(a_f + b_f)^{(a_f +
b_f)}}{{a_f}^{a_f} {b_f}^{b_f}} \; .
\label{admfct}
\ee
Here $a_f, b_f, N_f$ are free parameters obtained by fitting to data. Echevarria \emph{et al.} have obtained these parameters by a global fit of Sivers asymmetry in SIDIS using data on kaons, pions and charged hadrons from JLab, HERMES and COMPASS. These parameters are known for $u$ and $d$ quarks but there is no information available on $N_g,  a_g$ and $b_g$ for gluons.
In our work, we have used two parametrizations first proposed by Boer and Vogelsang~\cite{Boer-PRD69(2004)094025}:
\begin{align}
\text{(a)  }&{\mathcal N}_g(x)=\left( {\mathcal N}_u(x)+
{\mathcal N}_d(x) \right)/2 \;\nonumber\\
\text{(b)  }&{\mathcal N}_g(x)={\mathcal N}_d(x)
\label{sivers_parametrization}
\end{align}

In the first choice, it is assumed that the gluon Sivers fuction is the average of the up and down quark Sivers functions. The second choice is based on the fact that the gluon PDF is similar to the down quark PDF. Since these choices are essentially based on certain heuristic arguments, we explore the dependence of the asymmetry on these choices by comparing the estimates obtained with both.

The latest fits performed in Ref.~\cite{Echevarria:2014xaa} give the following values for the parameters of the quark Sivers function and the widths of the TMDs. This set was fitted at $Q_0 = \sqrt{2.4}\text{ GeV}$. We call this set TMD-e2,
\begin{gather}
 N_u = 0.106, \ a_u=1.051, \ b_u =4.857 \;, \nonumber \\
 N_d = -0.163, \ a_d = 1.552, \ b_d=4.857, \ b_{\text{max}}=1.5\text{ GeV}^{-1} \;, \nonumber \\
\langle k^2_{s\perp}\rangle = 0.282 \text{ GeV}^2, \ \langle k^2_{\perp}\rangle = 0.38\text{ GeV}^2, \ g_2 = 0.16 \text{ GeV}^2 
\label{exact2}
\end{gather}

These fits are used with the evolution formulation given above, in which the perturbative part is resummed upto NLL accuracy. 
Next, we include another set of parameters which we use to estimate the asymmetry, using the formulation provided in Ref.~\cite{Anselmino:2012aa}. These parameters, extracted at $Q_0 = 1.0\text{ GeV}$, are for the exact solution of TMD evolution equations. We call this set TMD-e1,

\begin{gather}
N_u = 0.77, \ a_u=.68, \ b_u =3.1  \;, \nonumber \\
N_d = -1.00, \ a_d = 1.11, \ b_d=3.1, \ b_{\text{max}}=0.5\text{ GeV}^{-1} \;, \nonumber \\
M_1^2 = 0.40 \text{ GeV}^2, \ \langle k^2_{\perp}\rangle = 0.25\text{ GeV}^2, \ g_2 = 0.68 \text{ GeV}^2 
\label{exact1}
\end{gather}
The asymmetries obtained with both parametrizations above are also compared with the results of Ref.~\cite{Godbole:2013bca} which is our previous paper where we have used an analytical solution of approximated TMD evolution equations. This analytical solution was obtained under the approximation that the perturbative part of the evolution kernel is independent of $b_T$ \cite{Anselmino:2012aa}. It should be noted that the parameter set for the analytical form is  different from the one in Eqn. (\ref{exact1}). We call this 
TMD-a set in Figures. (\ref{jlab_a})-(\ref{erhic2_b}) and give the parameters  below for completeness.
\begin{gather}
N_u = 0.75, \ a_u=.82, \ b_u =4.0  \;, \nonumber \\
N_d = -1.00, \ a_d = 1.36, \ b_d=4.0, \  b_{\text{max}}=0.5\text{ GeV}^{-1} \;, \nonumber \\
M_1^2 = 0..34 \text{ GeV}^2, \ \langle k^2_{\perp}\rangle = 0.25\text{ GeV}^2, \ g_2 = 0.68 \text{ GeV}^2 
\label{analytical}
\end{gather}

In all these cases $\langle k^2_\perp\rangle$ is given for quarks. We use the same value for gluons.
It should be noted that in the case of TMD-e1 and TMD-a fits, the appropriate Sivers function given in Ref. ~\cite{Anselmino:2012aa} was used.

For the asymmetry estimated using TMD-e2, MSTW2008LO gluon distribution was used. In the other cases, GRV98LO gluon distribution was used. These were the respective densities used by the authors of the different fits.  The TMDs were evaluated at $Q^2=\hat{s}$, which varies from $4m_c^2$ to $4m_D^2$. Here $m_c$ is the Charm quark mass, taken to be $1.275$ GeV and $m_D=1.863$ is the ``open-charm" threshold i.e., the D-meson mass.

The centre of mass(c.o.m) energies at which the asymmetry has been estimated are as follows: $\sqrt{s} = 4.7$ GeV (JLAB), 7.2 GeV (HERMES), 17.33 GeV (COMPASS), 31.6 GeV (eRHIC-1), 158.1 GeV (eRHIC-2).

 We would like to point out that  in our previous work, we had  calculated the kernel using the quark anomalous dimensions instead of gluon anomalous dimensions. However as the kernel, which is independent of $b_T$(in the analytical approximation) cancels between the numerator and the denominator of the asymmetry, this did not affect the result. In all the exact calculations here however, we have used gluon anomalous dimenions while calculating the kernel, as one should. For completeness, in Figure 1(a) we show a plot comparing the gluon Sivers function obtained using the quark anomalous dimensions and gluon anomalous dimensions. We note here that even in the case where we do not neglect the $b_T$ dependence of the kernel, the difference in the asymmetries obtained using a gluon kernel and a quark kernel are very small. This is because the kernel, which is in convolution with other factors is present both in the numerator and the denominator of the asymmetry and cancels out to a high degree.

\begin{figure}
\begin{center}
\includegraphics[width=0.45\linewidth,angle=0]{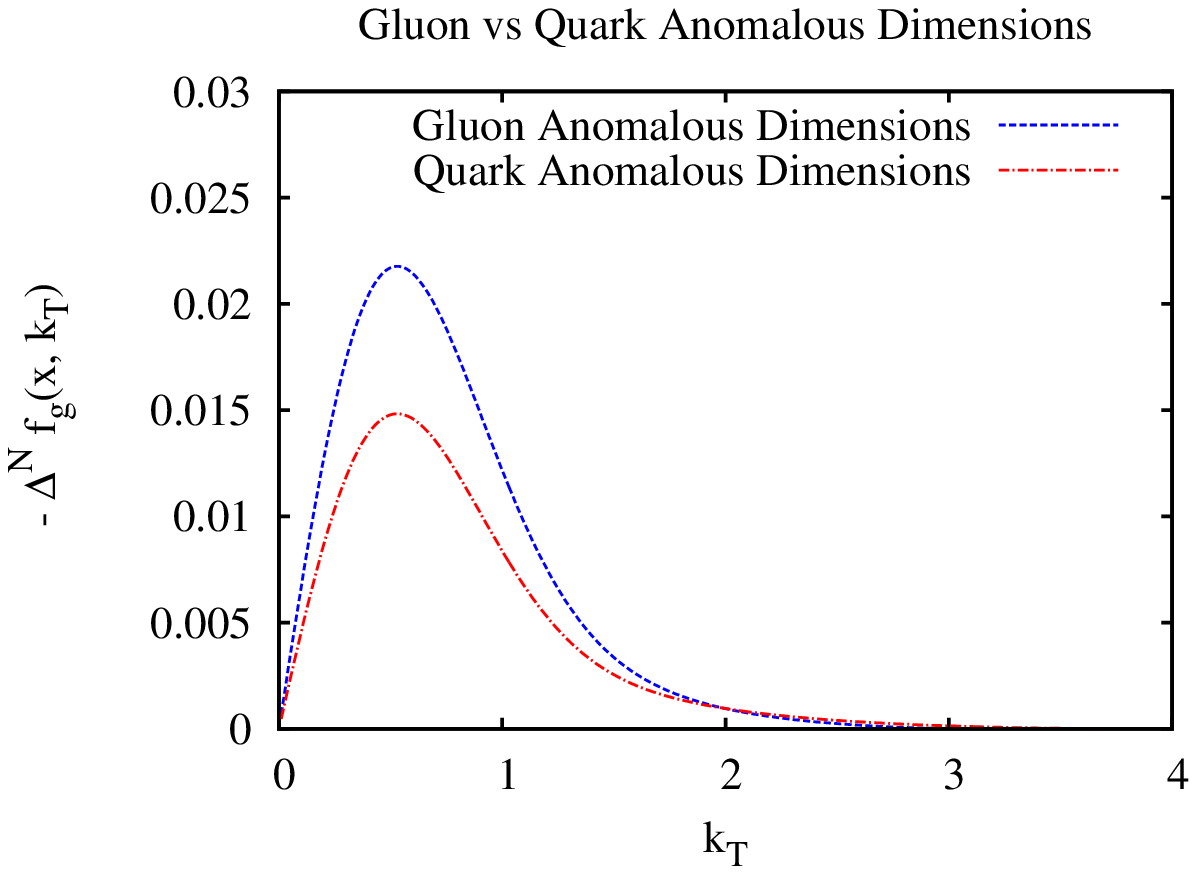}
\includegraphics[width=0.45\linewidth,angle=0]{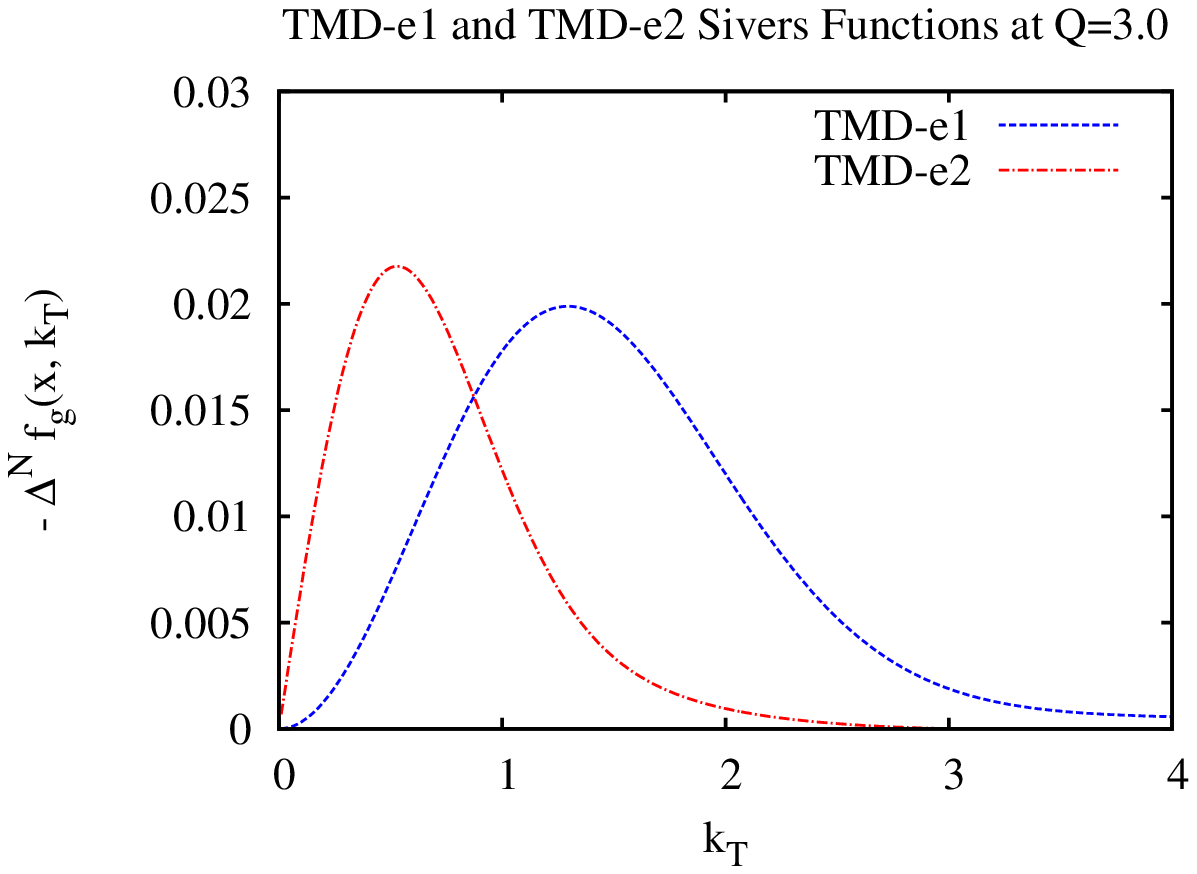}
\caption{Left-(a) Plot of the gluon Sivers functions in the TMD-e1 at $Q=3.0$ obtained using gluon and quark anomalous dimensions respectively, in the evolution kernel. Right-(b) Plot of the gluon Sivers functions at Q=3.0 obtained two using different fits- TMD-e1 and TMD-e2. The shift in the position of the peak has implications for the $q_T$ distribution of the Sivers asymmetry.}
\label{sivers_function}
\end{center}
\end{figure}

In Figure \ref{sivers_function}(b), we show the gluon Sivers function obtained using the fits TMD-e1 and TMD-e2. The Sivers function is plotted against $k_\perp$ at $Q=3.0$ GeV. The plot clearly shows that the peak of the Sivers function in the TMD-e1 fit occurs at a higher value of $k_T$ than in the TMD-e2 fit. This has implications for the $q_T$ distribution of the Sivers asymmetry. 

Next, we show the Sivers asymmetries obtained at JLab, HERMES, COMPASS , eRHIC-1 and eRHIC-2 energies for both parametrizations (a) and (b) of the Sivers function shown in Eqn. (\ref{sivers_parametrization}). The asymmetry as a function of $y$ was obtained by integrating Eqn. (\ref{weight-ssa}) over $q_T$ (from 0 GeV to 1 GeV). We call this the ``$y$-asymmetry". The asymmetry as a function of $q_T$ was obtained by integrating Eqn. (\ref{weight-ssa}) over the kinematically allowed ranges of $y$. These were: $-0.25\leq y\leq0.25$ for JLab, $-0.6\leq y\leq 0.6$ for HERMES, $-1.5\leq y\leq1.5$ for COMPASS, $-2.1\leq y\leq2.1$ for eRHIC-1 and $-3.7\leq y\leq3.7$ for eRHIC-2. We call this the ``$q_T$-asymmetry". It should be noted that the rapidity, $y$, given here is in the centre of mass frame of the colliding electron and proton.

Figures (\ref{jlab_a}) to (\ref{erhic2_a}) show the asymmetries as a function of $y$ and $q_T$ with increasing center of mass energies from JLab (4.7 GeV) to e-RHIC (158.1 GeV). We do this for all three TMD evolved fits as well as the DGLAP fit with parametrization (a) of the gluon Sivers function as given in Eqn. (\ref{sivers_parametrization}). Figures (\ref{jlab_b}) to (\ref{erhic2_b}) show the same but for parametrization (b) of the gluon Sivers function. 

\vfill
\begin{figure}
\begin{center}
\includegraphics[width=0.45\linewidth,angle=0]{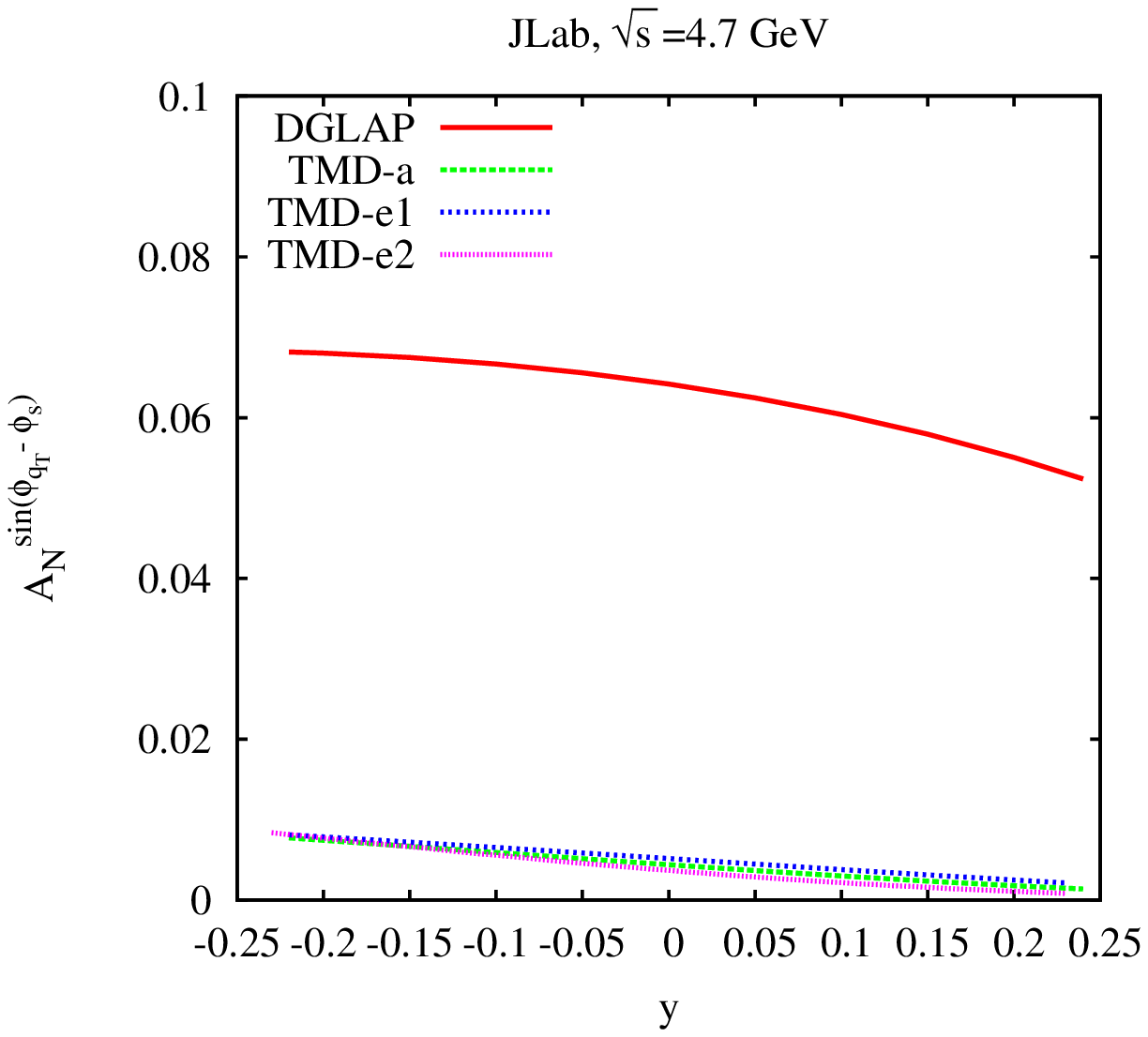}
\includegraphics[width=0.45\linewidth,angle=0]{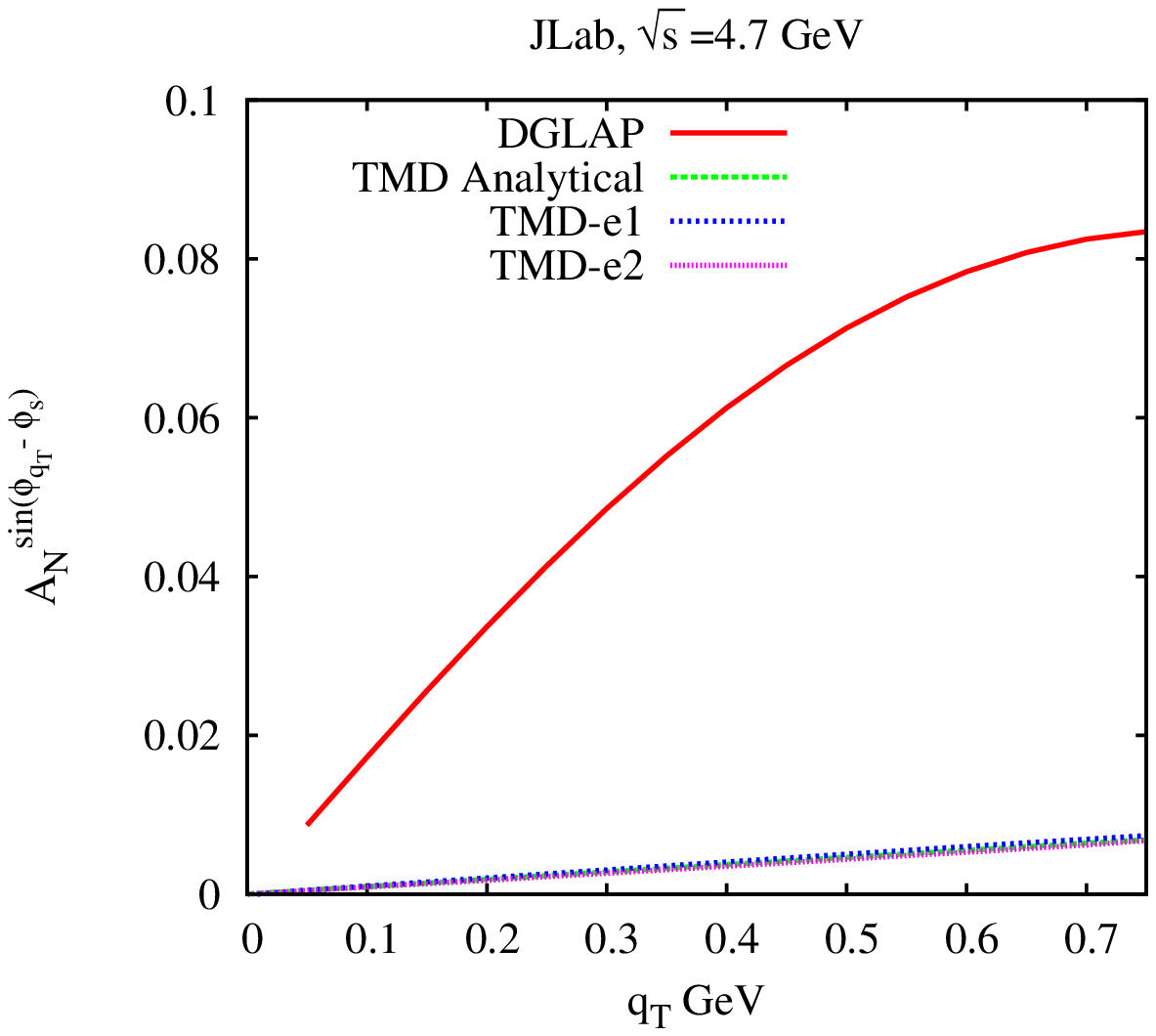}
\caption{The Sivers asymmetry $A_N^{\sin({\phi}_{q_T}-\phi_S)}$ for $e+p^\uparrow \to  e+J/\psi +X $
at JLab energy ($\sqrt{s} = 4.7$~GeV), as a function of $y$ (left panel) and $q_T$ (right panel)
for {\it parameterization (a)}. The dotted green line corresponds to the TMD analytical approximated evolution approach given in Ref.~\cite{Godbole:2013bca} and dotted blue line corresponds to exact TMD evolution approach with TMD-e1 parameter set. The dotted pink line corresponds to the TMD evolution results using the CSS approach with TMD-e2 parameter set and the solid red line corresponds to the asymmetry obtained using the DGLAP evolution given in Ref.~\cite{Godbole:2012bx}.
The integration ranges are $(0 \leq q_T \leq 1)$~GeV and $(-0.25 \leq y \leq 0.25)$.}
\label{jlab_a}
\end{center}
\end{figure}
\begin{figure}
\begin{center}
\includegraphics[width=0.45\linewidth,angle=0]{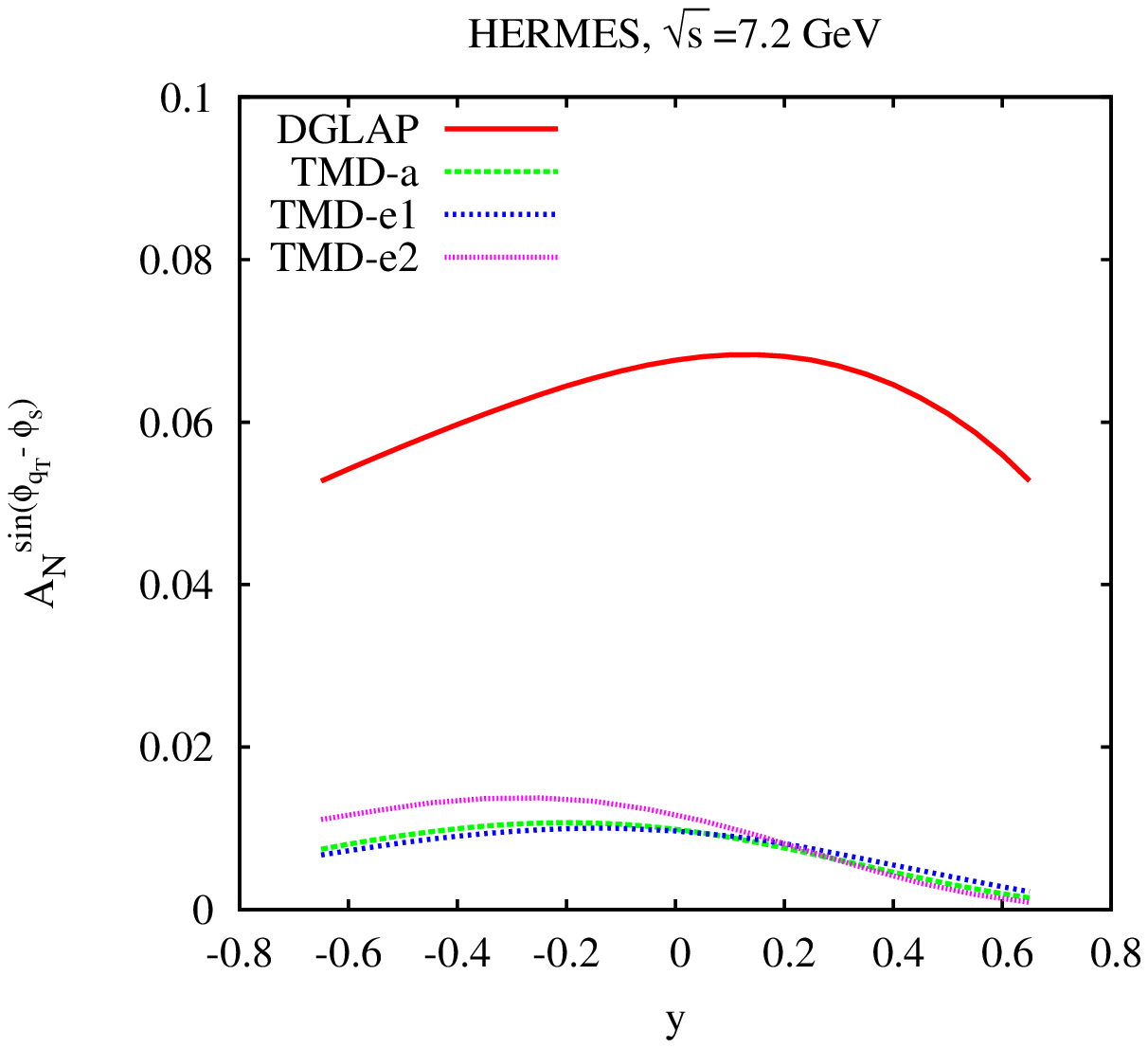}
\includegraphics[width=0.45\linewidth,angle=0]{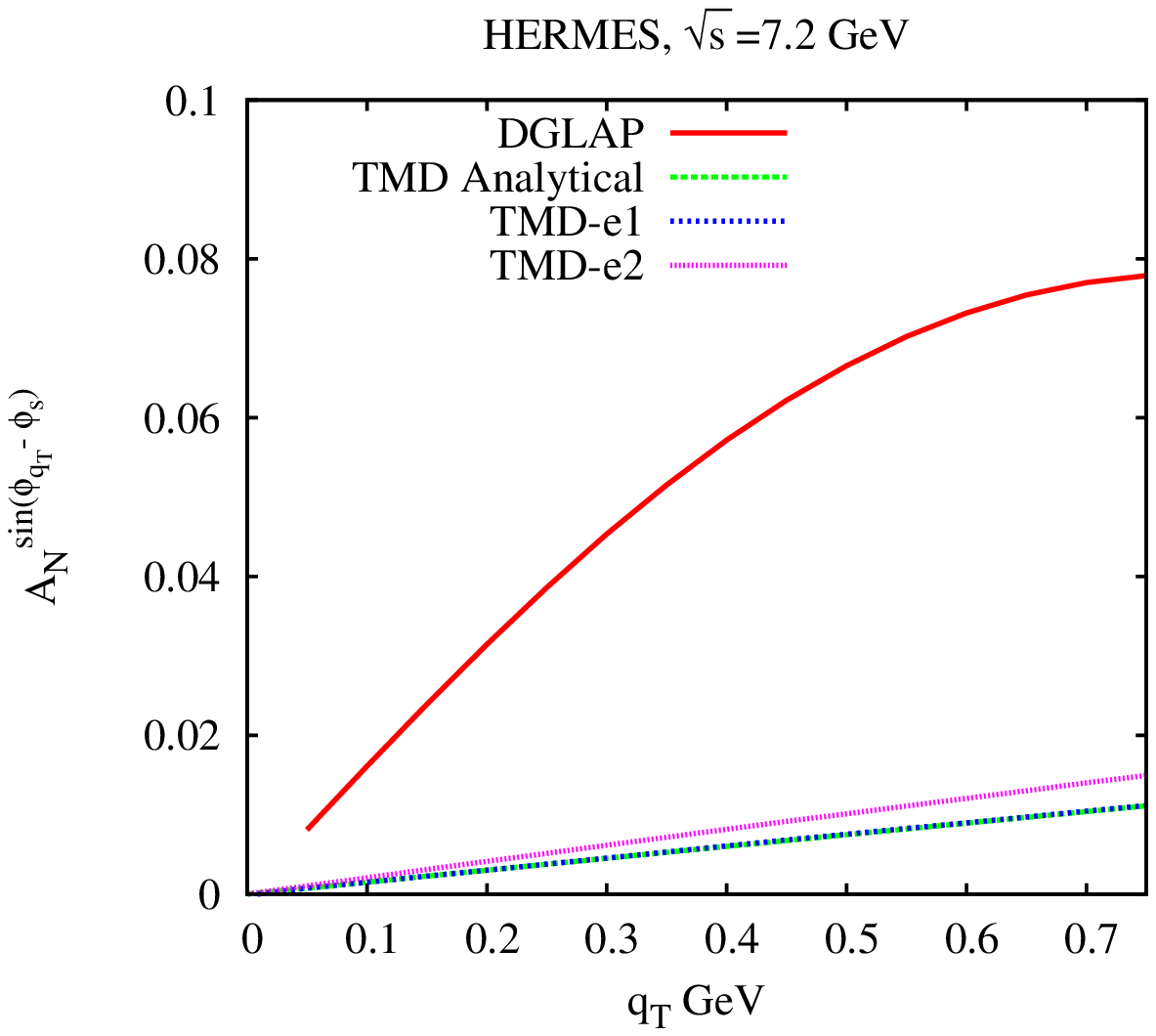}
\caption{ HERMES energy ($\sqrt{s} = 7.2$~GeV), Asymmetry as a function of $y$ (left panel) and $q_T$ (right panel)
for {\it parameterization (a)}. The integration ranges are $(0 \leq q_T \leq 1)$~GeV and $(-0.6 \leq y \leq 0.6)$. The convention for the colour and line styles is the same as in Figure 2.}
\label{hermes_a}
\end{center}
\end{figure}
\begin{figure}
\begin{center}
\includegraphics[width=0.45\linewidth,angle=0]{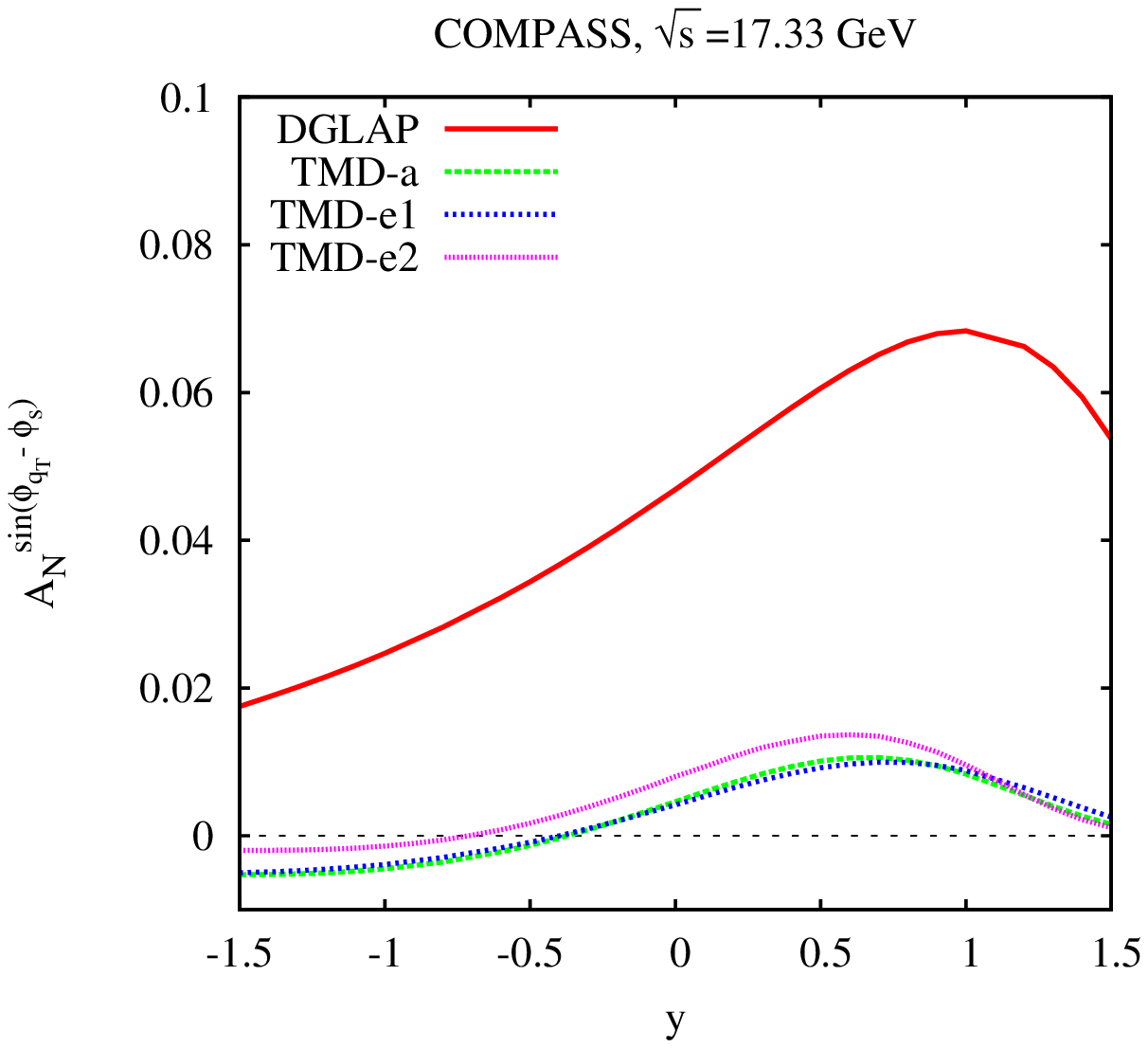}
\includegraphics[width=0.45\linewidth,angle=0]{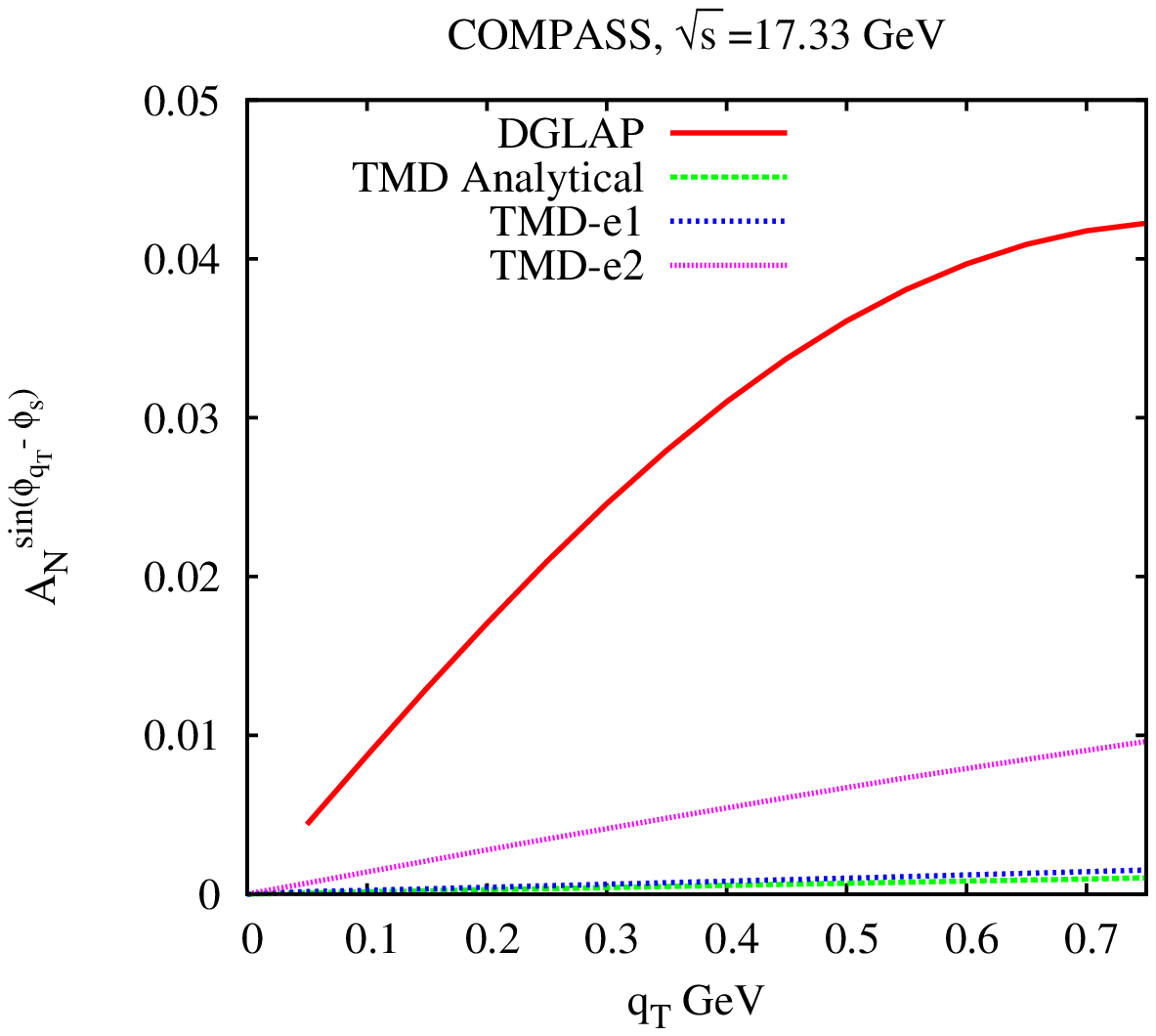}
\caption{COMPASS energy ($\sqrt{s} = 17.33$~GeV), Asymmetry as a function of $y$ (left panel) and $q_T$ (right panel) for {\it parameterization (a)}. The integration ranges are $(0 \leq q_T \leq 1)$~GeV and $(-1.5 \leq y \leq 1.5)$. The convention for the colour and line styles is the same as in Figure 2.}
\label{compass_a}
\includegraphics[width=0.45\linewidth,angle=0]{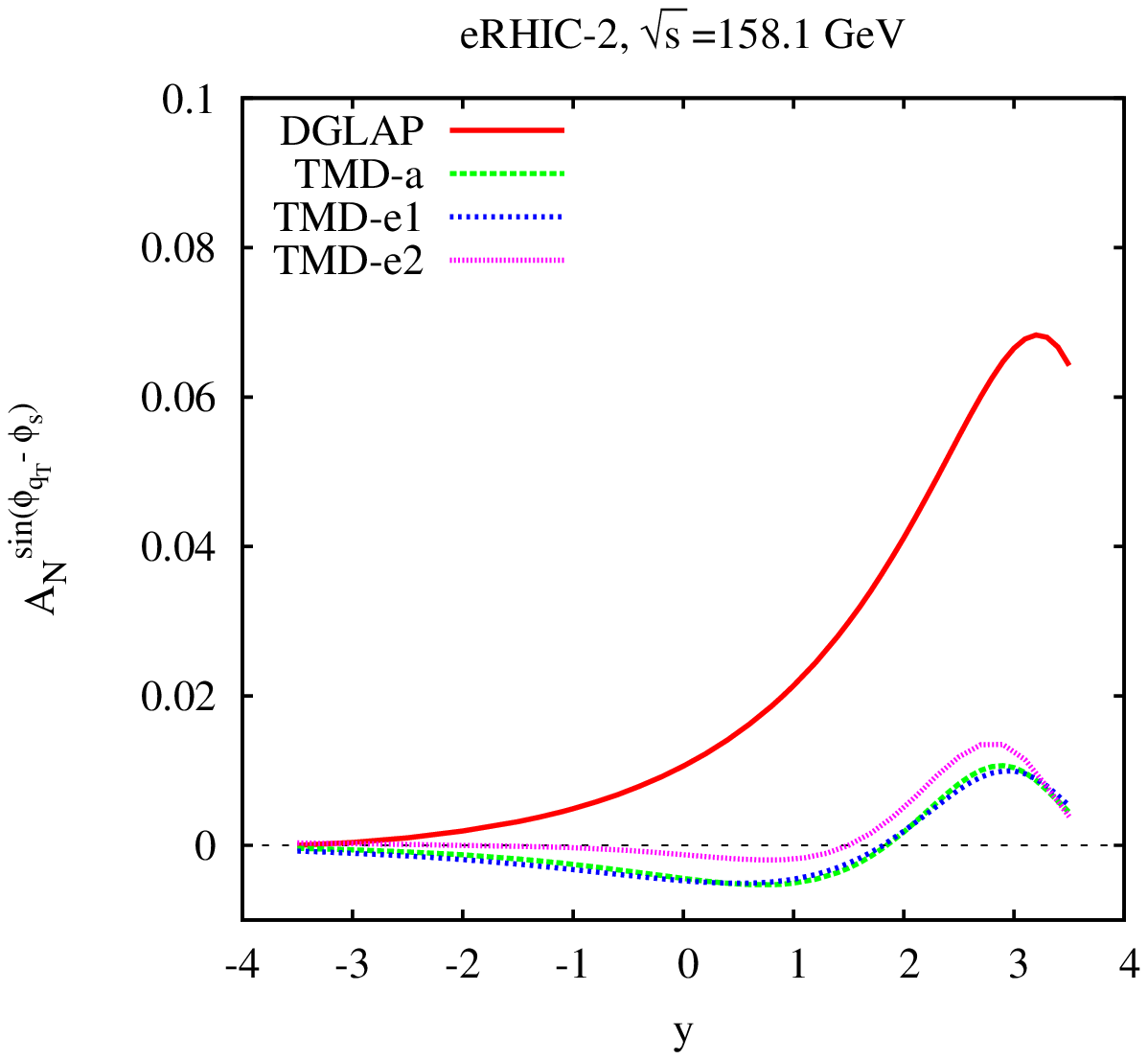}
\includegraphics[width=0.45\linewidth,angle=0]{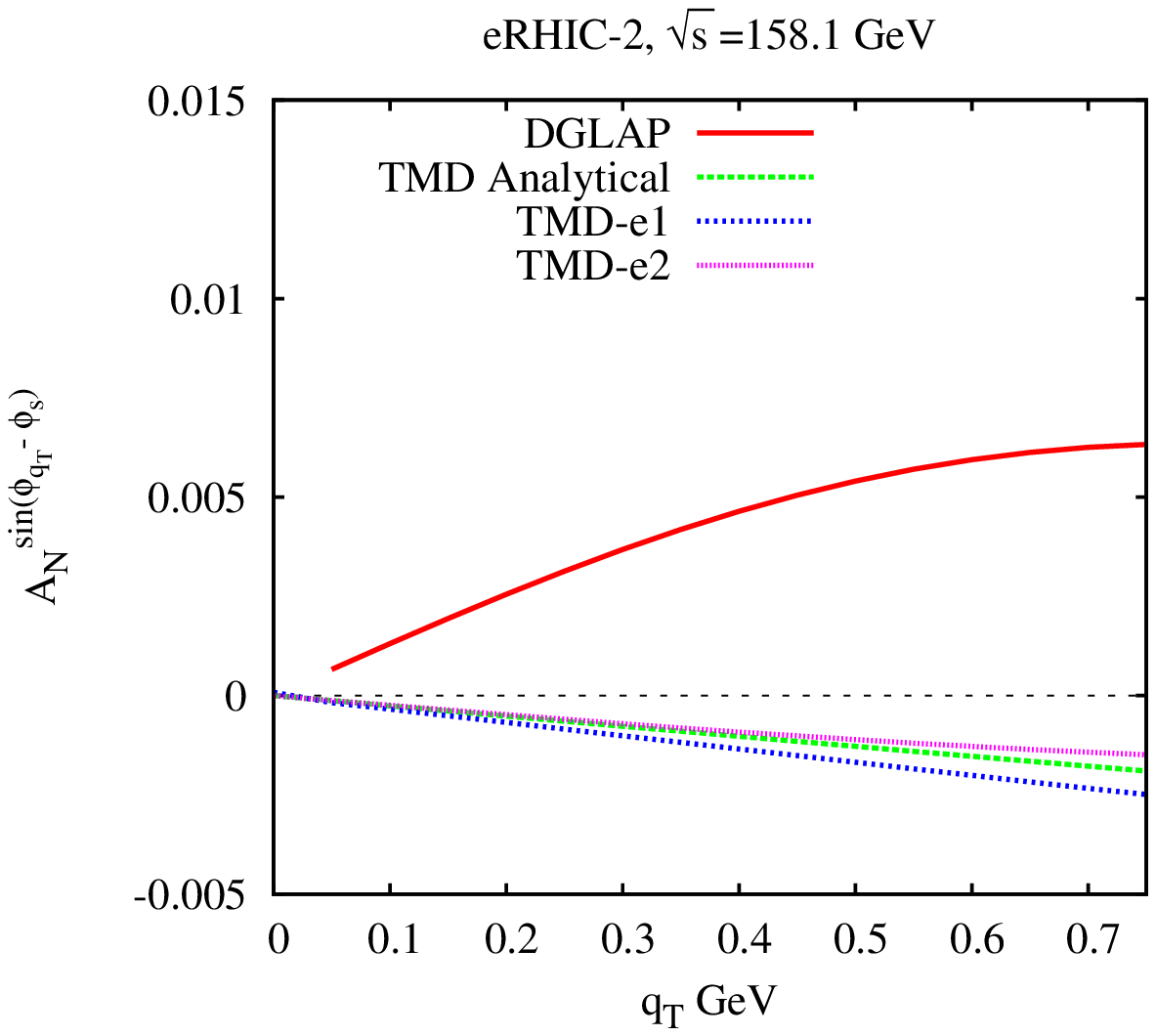}
\caption{eRHIC energy ($\sqrt{s} = 158.1$~GeV),Asymmetry as a function of $y$ (left panel) and $q_T$ (right panel)
for {\it parameterization (a)}. The integration ranges are $(0 \leq q_T \leq 1)$~GeV and $(-3.7 \leq y \leq 3.7)$. The convention for the colour and line styles is the same as in Figure 2.}
\label{erhic2_a}
\end{center}
\end{figure}
\begin{figure}
\begin{center}
\includegraphics[width=0.45\linewidth,angle=0]{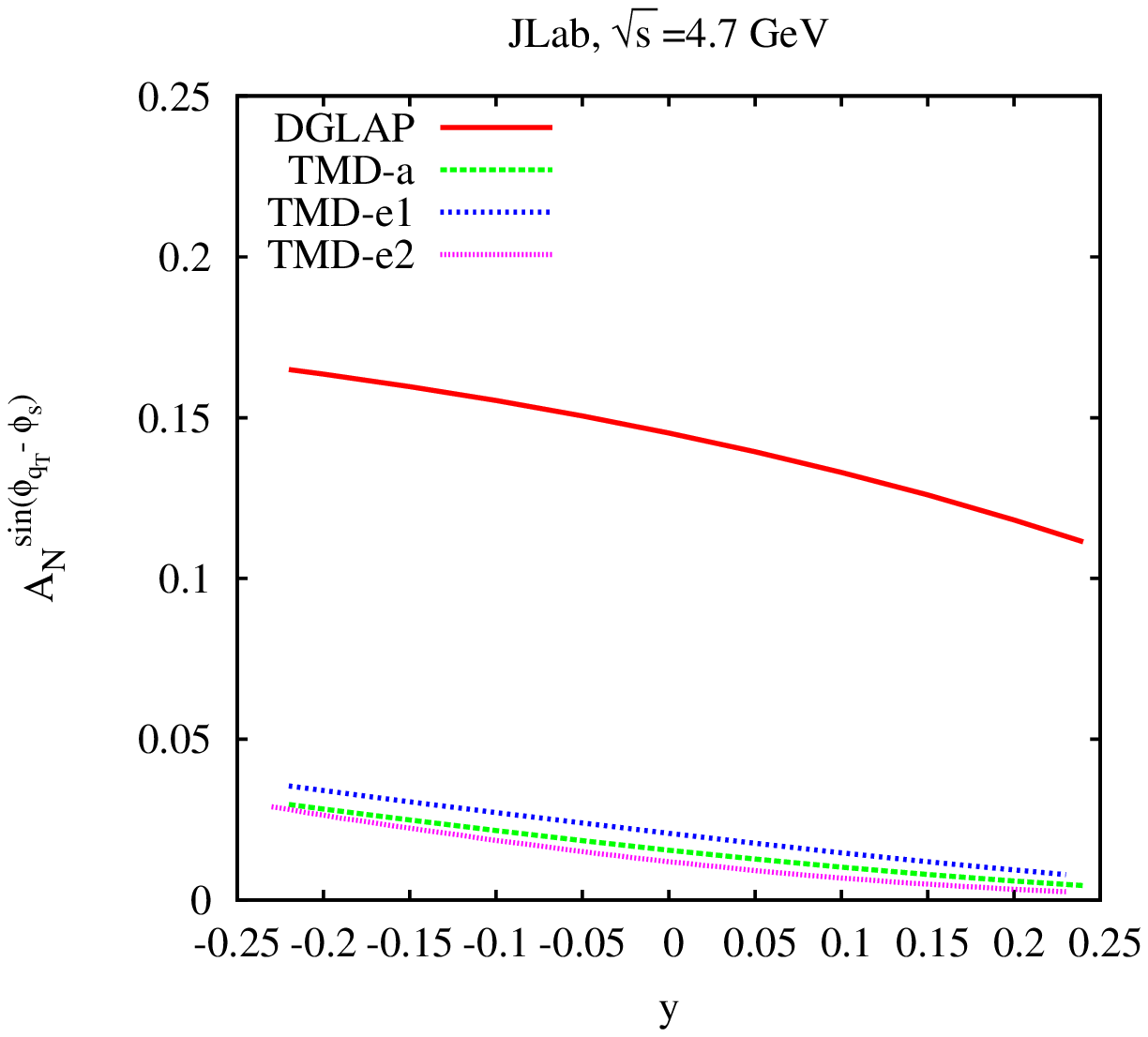}
\includegraphics[width=0.45\linewidth,angle=0]{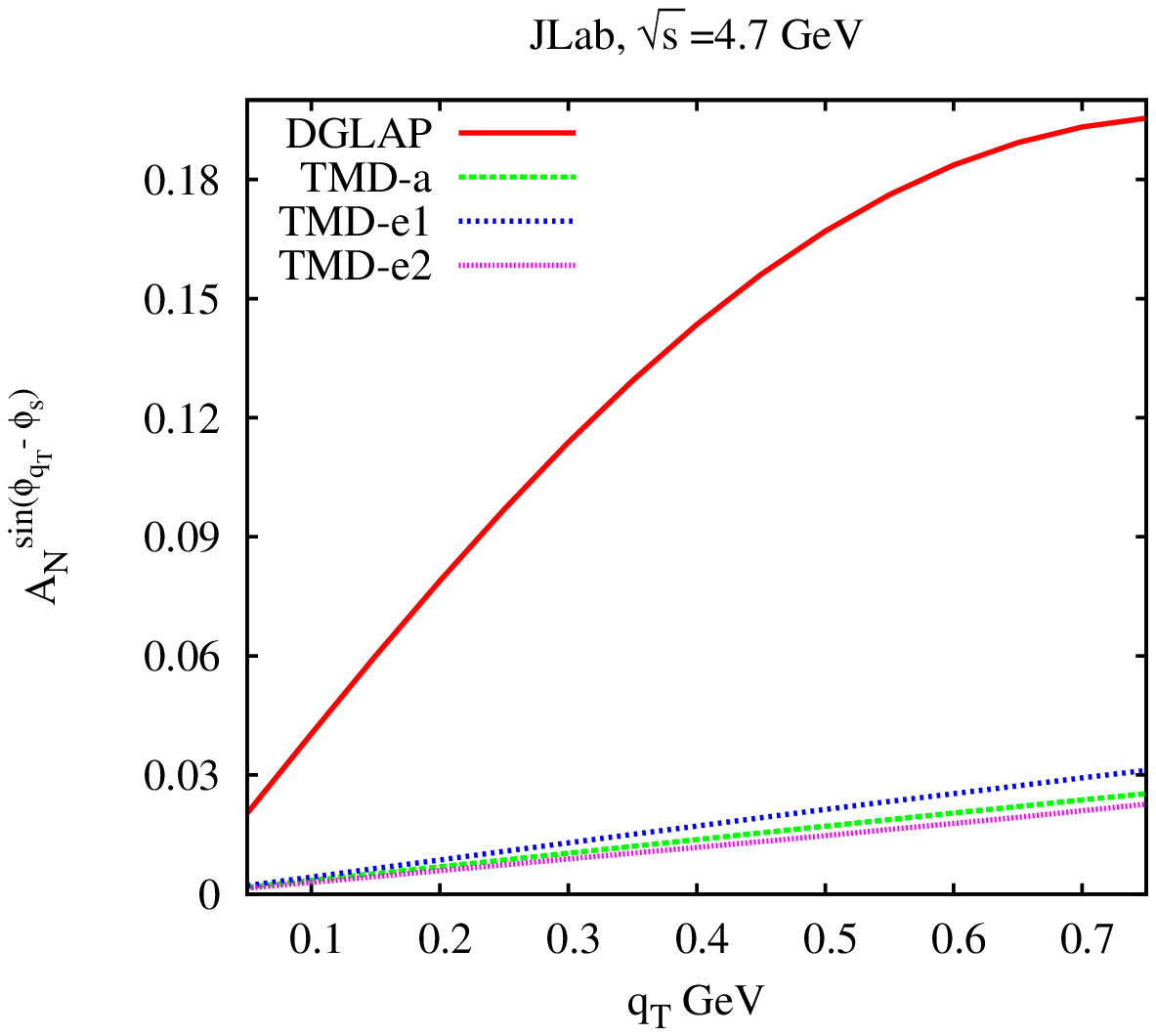}
\caption{JLab energy ($\sqrt{s} = 4.7$~GeV), Asymmetry as a function of $y$ (left panel) and $q_T$ (right panel)
for {\it parameterization (b)}. The integration ranges are $(0 \leq q_T \leq 1)$~GeV and $(-0.25 \leq y \leq 0.25)$. The convention for the colour and line styles is the same as in Figure 2.}
\label{jlab_b}
\includegraphics[width=0.45\linewidth,angle=0]{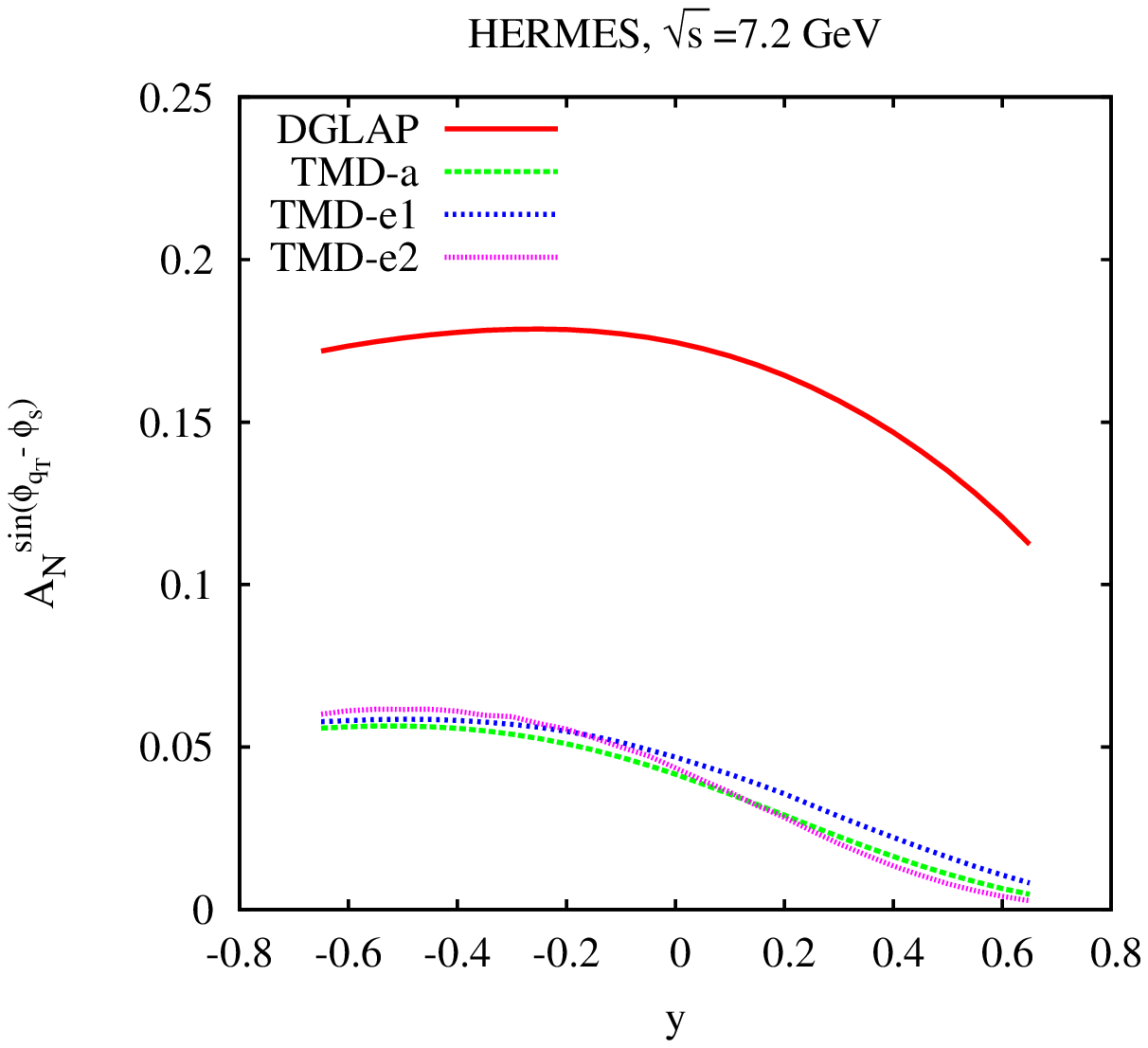}
\includegraphics[width=0.45\linewidth,angle=0]{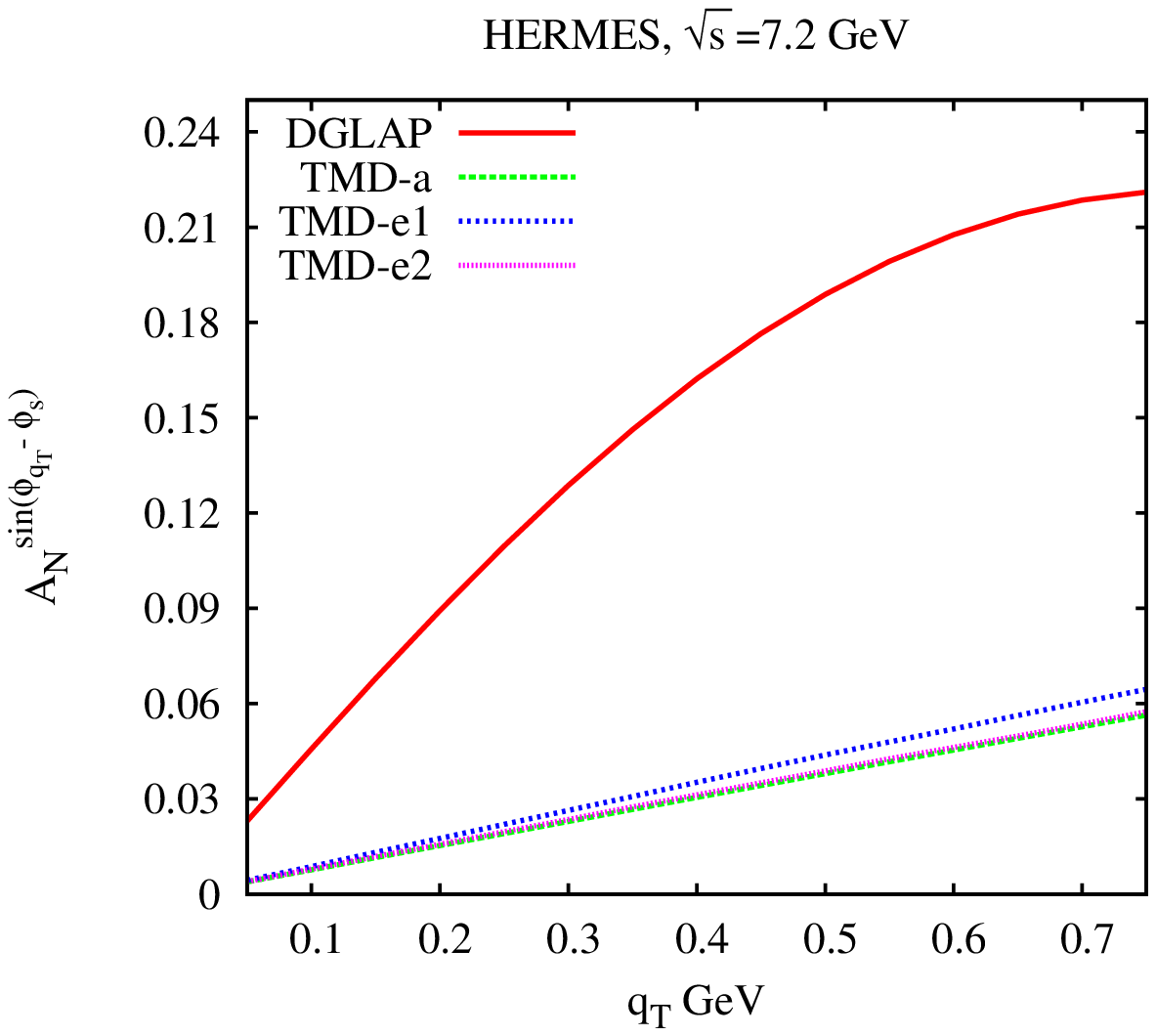}
\caption{HERMES energy ($\sqrt{s} = 7.2$~GeV), Asymmetry as a function of $y$ (left panel) and $q_T$ (right panel)
for {\it parameterization (b)}. The integration ranges are $(0 \leq q_T \leq 1)$~GeV and $(-0.6 \leq y \leq 0.6)$. The convention for the colour and line styles is the same as in Figure 2.}
\label{hermes_b}
\end{center}
\end{figure}
\begin{figure}
\begin{center}
\includegraphics[width=0.45\linewidth,angle=0]{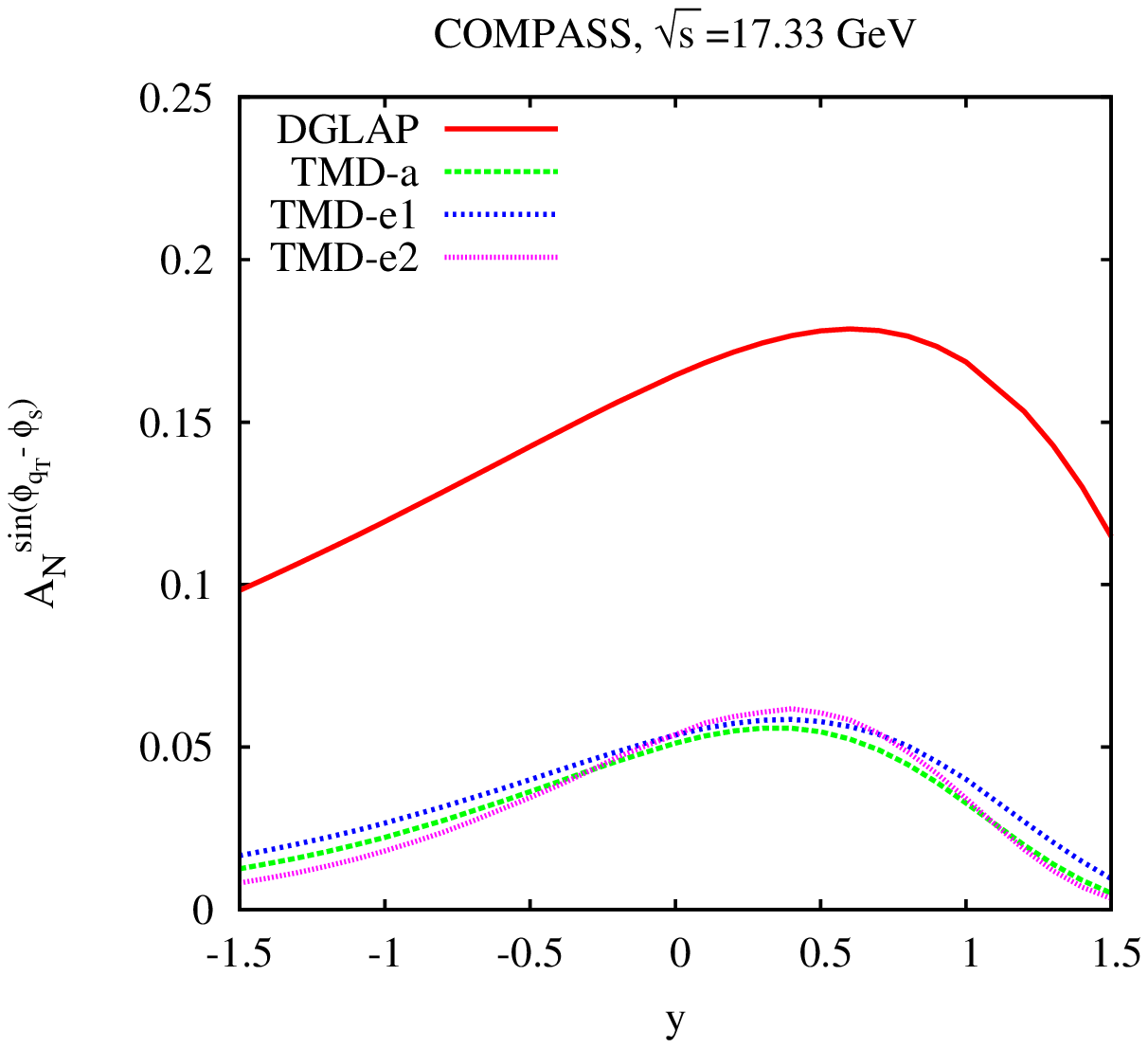}
\includegraphics[width=0.45\linewidth,angle=0]{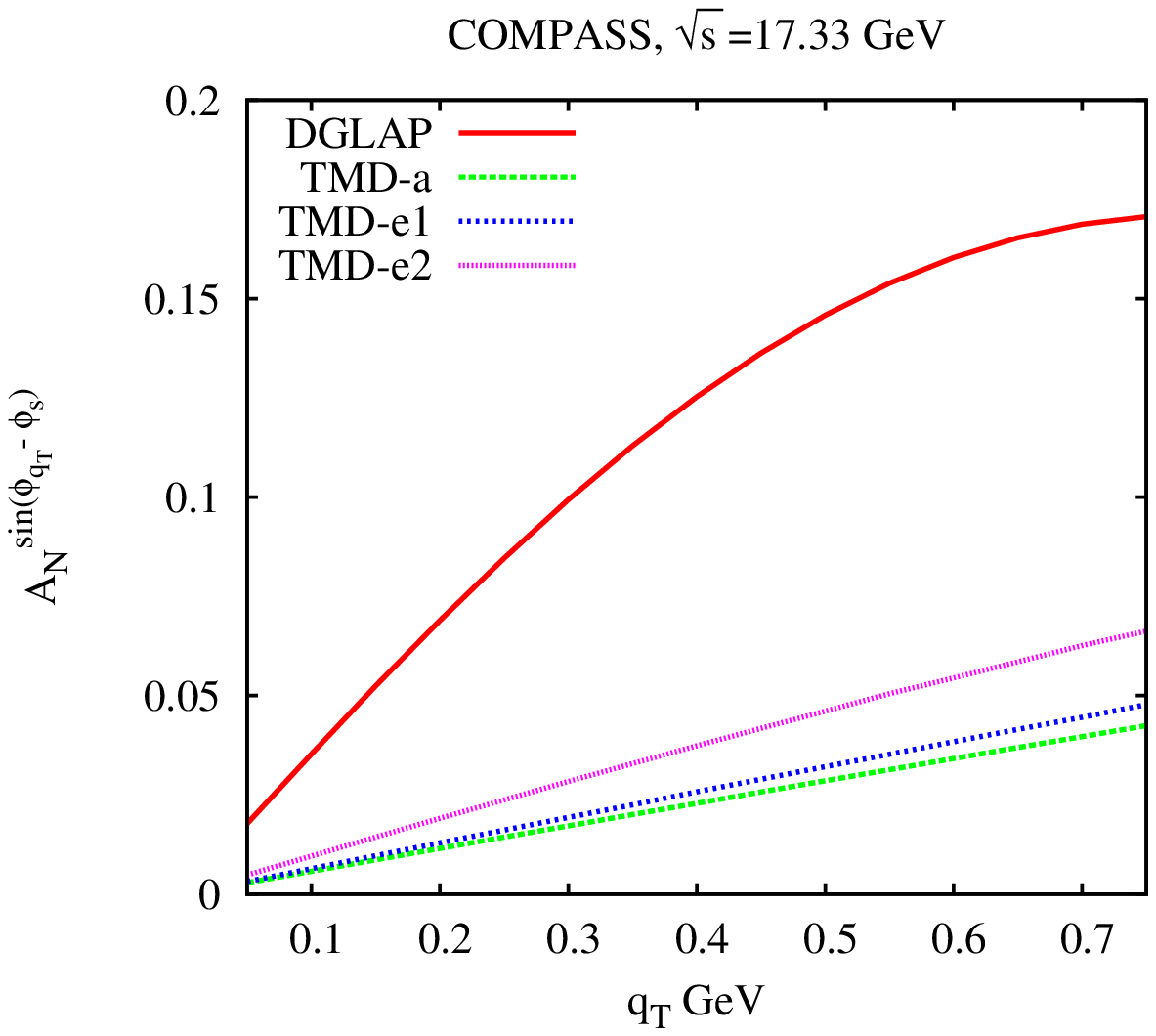}
\caption{COMPASS energy ($\sqrt{s} = 17.33$~GeV), Asymmetry as a function of $y$ (left panel) and $q_T$ (right panel)
for {\it parameterization (b)}. The integration ranges are $(0 \leq q_T \leq 1)$~GeV and $(-1.5 \leq y \leq 1.5)$. The convention for the colour and line styles is the same as in Figure 2.}
\label{compass_b}
\includegraphics[width=0.45\linewidth,angle=0]{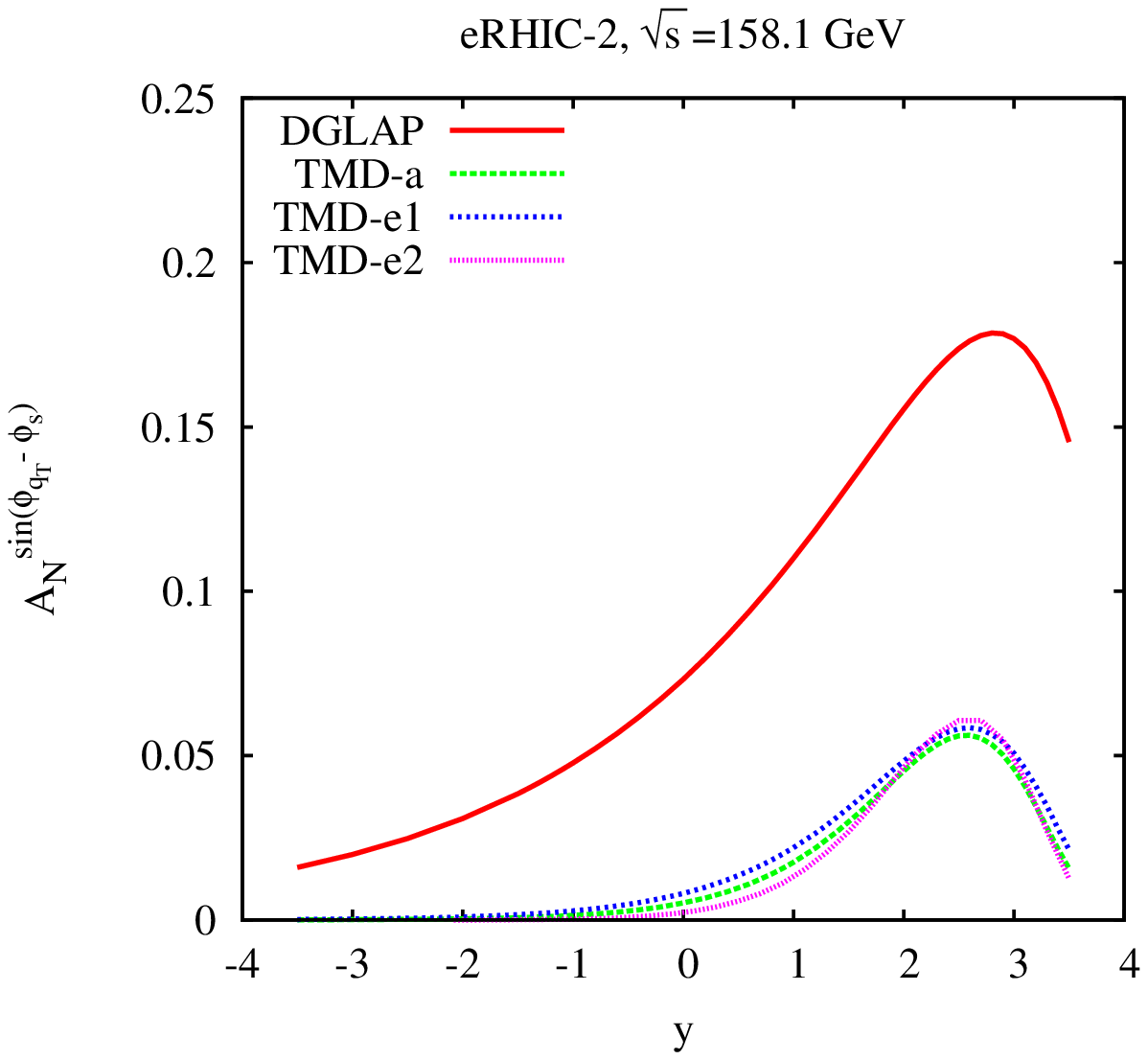}
\includegraphics[width=0.45\linewidth,angle=0]{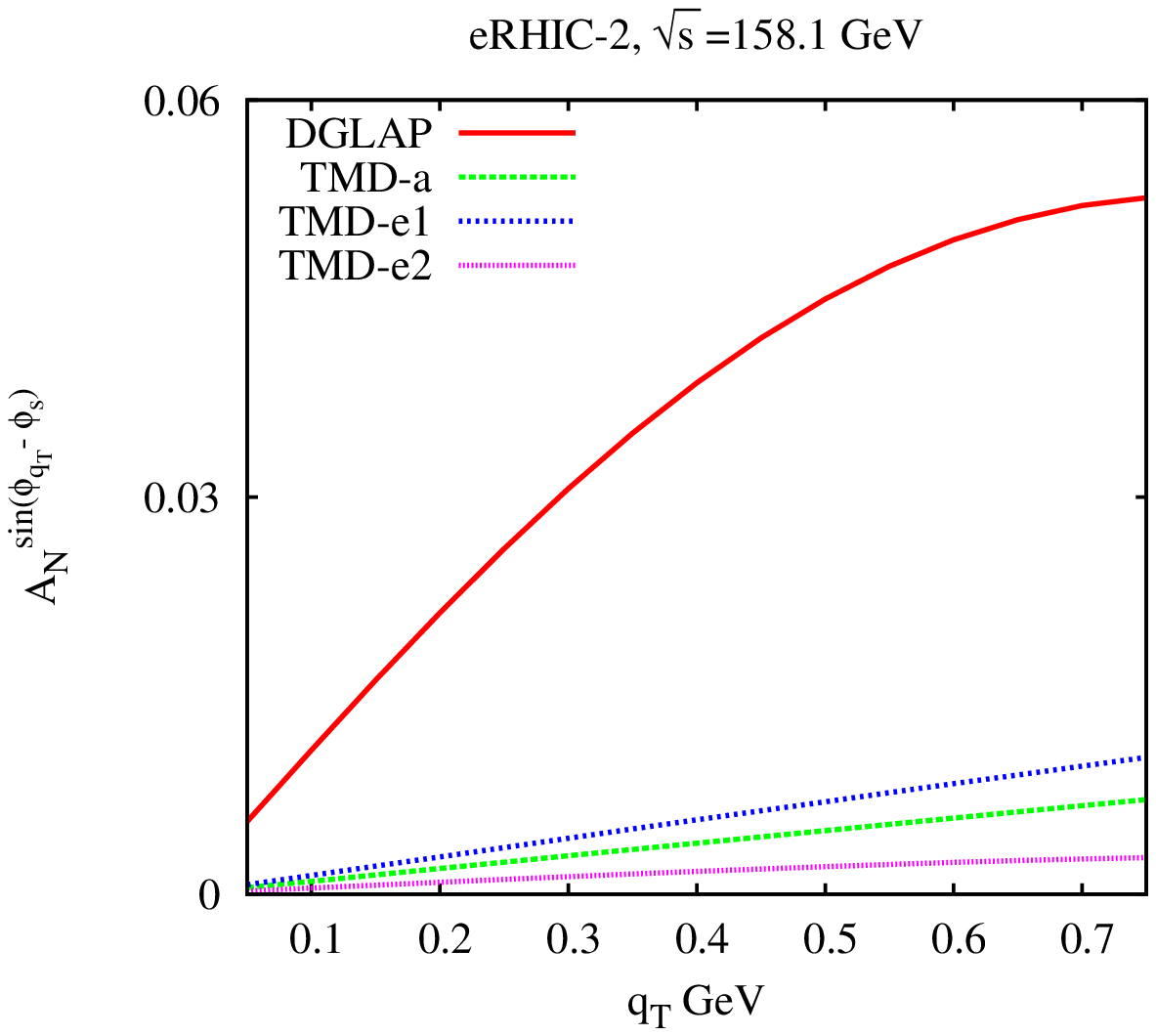}
\caption{eRHIC energy ($\sqrt{s} = 158.1$~GeV), Asymmetry as a function of $y$ (left panel) and $q_T$ (right panel)
for {\it parameterization (b)}. The integration ranges are $(0 \leq q_T \leq 1)$~GeV and $(-3.7 \leq y \leq 3.7)$. The convention for the colour and line styles is the same as in Figure 2.}
\label{erhic2_b}
\end{center}
\end{figure}

Now we discuss the results in detail. We are broadly concerned with two things: The behaviour of the predictions under different treatments of transverse momentum dependent evolution, and the dependence of the predictions on the center of mass energy $\sqrt{s}$.

First we look at the behaviour under different treatments of evolution. Figures (\ref{jlab_a})-(\ref{erhic2_a}) show the $y$- and $q_T$-asymmetry estimates for JLAB, HERMES, COMPASS and eRHIC-2 energies obtained using parametrization (a) of the gluon Sivers function. Figures (\ref{jlab_b})-(\ref{erhic2_b}) show estimates obtained with parametrization (b). We can see that the asymmetries given by the TMD-a and TMD-e1 fits are similar. This was to be expected as both use the same kernel (except for the approximation on the $b_T$ dependence) and were fitted to the same data. This shows that the approximation made in obtaining the analytical solution in Ref.~\cite{Anselmino:2012aa} is a good one.  The $y$-asymmetries given by the TMD-e2 fits are also similar in size to the asymmetries given by TMD-a and TMD-e1 for parametrization (b) of the gluon Sivers function. However, they are slightly larger in case of parametrization (a). We would like to remind here that the perturbative part of the kernel used with the TMD-e2 fit has been resummed at NLL~\cite{Echevarria:2014xaa}. The asymmetries obtained using all three TMD evolved fits are much smaller than those obtained using DGLAP evolution, but amongst themselves, they are similar in size. That is, the predictions are stable over different treatments of TMD evolution. This is the main result of our analysis.

As was the case with the $y$-asymmetries, the three TMD evolved predictions of the $q_T$-asymmetry are much smaller than the DGLAP prediction, but similar amongst themselves. In general, the $q_T$ asymmetries for different choices of evolution vary more amongst each other when compared to the $y$-asymmetries but again, the asymmetries obtained using TMD-a and TMD-e1 are similar.  The magnitude of the $q_T$ asymmetries increases monotonically with $q_T$ in the range considered ($0\leq q_T\leq 0.75$ GeV). For parametrization (a) of the gluon Sivers function they even become negative for eRHIC-2 energies, as can be seen in Figure (\ref{erhic2_a}). 

The minor differences that we observe in the behaviour of the $q_T$-asymmetries with different TMD fits can be understood in terms of the differences in the $k_T$ behaviour of the Sivers functions obtained with different fits. Figure 1(b) shows the Sivers functions obtained using the TMD-e1 and TMD-e2 fits for $Q=3.0$ GeV and $x=0.1$. Since the transverse momentum behaviour of the partons is directly reflected in the transverse momenta of the produced $J/\psi$ pairs, the $k_T$ behaviour of the Sivers function influences the $q_T$-asymmetry.

We note that there is a substantial difference in the magnitude of the asymmetries between the two parametrizations of the gluon Sivers function. The peak asymmetry in the $y$-distribution obtained using TMD-e2 fit varies between $1.3\%$ for parametrization (a) of the Sivers function and $6.1\%$ for parametrization (b) for all energies except JLab. This is because the kinematics at the c.o.m energy of $4.7$ GeV allows contributions only from a region where the Sivers function is small. One can see from Eqn. (\ref{x-gammag}) that only contributions in the region $x_g\geq 0.42$ are allowed, causing only the tail-end of the gluon Sivers function to affect the process.

\begin{figure}
\begin{center}
\includegraphics[width=.45\linewidth,angle=0]{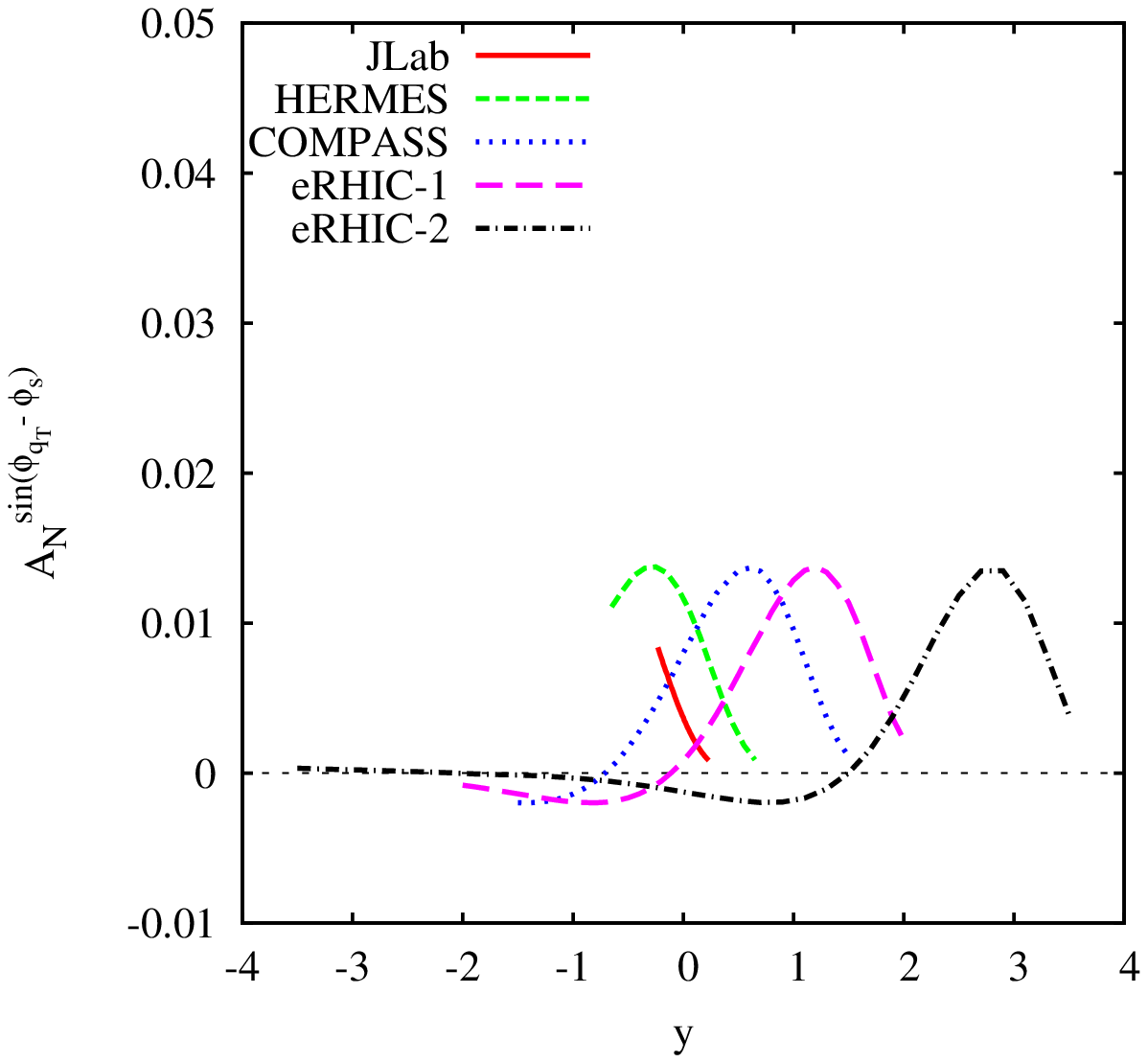}
\includegraphics[width=.45\linewidth,angle=0]{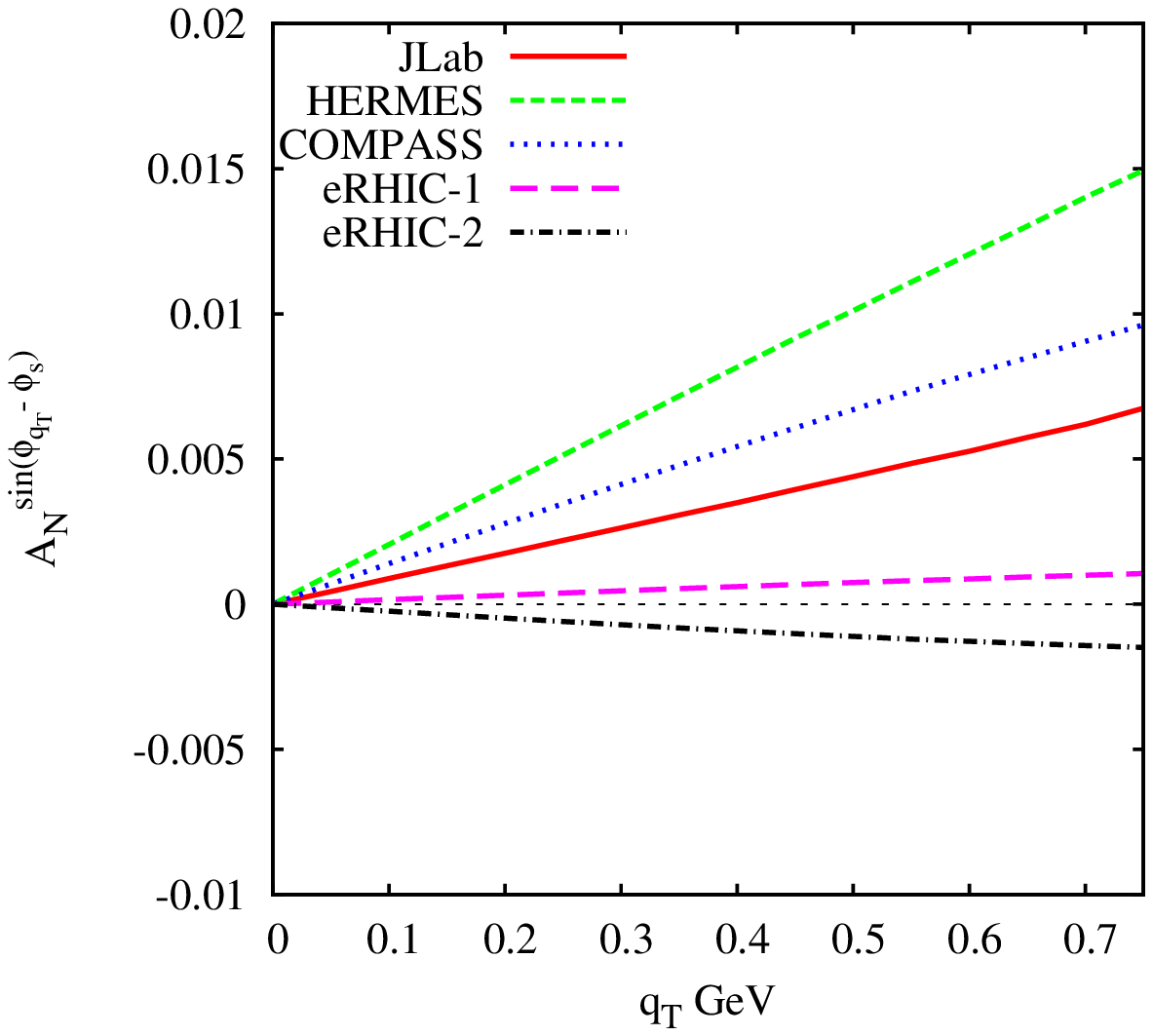}
\caption{Left panel: Plot of the Sivers asymmetries in the $y$ distribution obtained in all c.o.m energies using the TMD-e2 fit and parametrization (a) of the gluon Sivers function. This plot shows the drift of the asymmetry peak towards higer values of rapidity $y$. Right panel: Plot of the Sivers Asymmetries in the $q_T$ distribution. This shows the general decrease of asymmetry values with increasing c.o.m energy. The solid red line corresponds to the asymmetry at JLAB($\sqrt{s}=4.7$) energy, the dashed green line corresponds to HERMES($\sqrt{s}=7.2$) energy, the dotted blue line corresponds to COMPASS($\sqrt{s}=17.33$) energy, the dashed pink line corresponds to eRHIC-1($\sqrt{s}=31.6$) energy and the dashed-dotted black line line corresponds to eRHIC-2($\sqrt{s}=158.1$) energy.}
\label{compare_y}
\end{center}
\end{figure}
Now we look at the dependence of the asymmetry estimates on the center of mass energy $\sqrt{s}$. Figure (\ref{compare_y}) shows the estimates obtained using TMD-e2 set and parametrization (a) of the gluon Sivers function, for all energies. The $y$ and $q_T$ asymmetries for all considered values of $\sqrt{s}$ for the TMD-e2 fit are shown in Figure (\ref{compare_y}). In the $y$-asymmetry, we note that the peak shifts to higher values of $y$ with increasing $\sqrt{s}$. This shift can be understood when one notes that the position of the peak asymmetry has a strong dependence on the gluon momentum fraction $x_g$ through the Sivers function. The momentum fraction $x_g$ is directly related to rapidity through Eqn. (\ref{x-gammag}). From Eqn. (\ref{x-gammag}), we see that for increasing $\sqrt{s}$, the
 value of $y$ at which the Sivers function peaks, increases. Therefore the peak shifts towards higher values of $y$. More generally, the $y$-asymmetry gives us information on the $x_g$ dependence of the Sivers function; increasing the c.o.m energy allows us to probe larger ranges of $y$ and hence, larger ranges of $x_g$.

Apart from JLab, the asymmetry predictions in the $q_T$ distribution become smaller with increasing $\sqrt{s}$.  As pointed out earlier, In the case of eRHIC energies and parametrization (a), the asymmetries even become negative. The asymmetry at JLAB energy does not conform to this trend.  Again, as with the $y$ distribution, this is due to the constraint on the gluon momentum fraction $x_g$. 

We note that the $q_T$-asymmetries given here are for the full allowed kinematic ranges of $y$. We have not considered the acceptances of different experiments. It would be interesting to see how the predictions are affected when the details of the experiments are considered.
\vfill

\section{SUMMARY AND CONCLUSION}
The single spin asymmetry in the low virtuality electroproduction of $J/\psi$ has been estimated with respect to its transverse momentum and its rapidity using a NLL-resummed TMD evolution formulation.  The results for JLAB, COMPASS, HERMES, and eRHIC energies are presented at the end. It is observed that the estimates obtained are much smaller than those obtained earlier using DGLAP evolution with a non-evolving gaussian form for the unpolarized PDF and the Sivers function. The asymmetries calculated here are also similar in size to those obtained using an earlier fit(TMD-e1) to the SIDIS data wherein the TMD evolution formalism does not include the NLL resummation. We further observe that in the case of TMD-e1 fits, the difference between these asymmetries and those reported by us earlier, calculated using fits to SIDIS data based on an approximate, analytical form of the TMD evolution equation, is small. To summarize therefore, the asymmetries obtained using the TMD evolution are consistently much smller than those without it and further use of the TMD evolved Sivers function stabilizes the predictions for the size of the asymmetries. Overall the asymmetries remain sizable.

\section{Acknowledgement}

We are grateful to Dr. Zhong-Bo Kang for pointing out an inconsistency in the procedure used by us. R.M.G. wishes to acknowledge support from the Department of Science and
Technology, India under Grant No. SR/S2/JCB-64/2007 under the J.C. Bose 
Fellowship scheme. A.M. would like to thank the Indian National Science Academy for a travel fellowship and 
CHEP, IISc Bangalore for their kind hospitality. A.M and V.S.R would like to thank the Department of Atomic Energy-BRNS, India, for financial support under the grant No. 2010/37P/47/BRNS.

\pagebreak

\end{document}